\documentclass[apj]{emulateapj}

\usepackage{lscape}  

\newcommand{\etal}{et~al.}

\newcommand{\cgsflux}{erg~s$^{-1}$~cm$^{-2}$}
\newcommand{\ergss}{erg~s$^{-1}$}
\newcommand{\cmsq}{\hbox{cm$^{-2}$}}
\newcommand{\OIIlong}{{\rm O}\kern 0.1em{\sc ii}~$\lambda 3727$} 
\newcommand{\OII}{{\rm O}\kern 0.1em{\sc ii}} 
\newcommand{\Lsun}{L$_{\odot}$}
\newcommand{\spitzer}{\emph{Spitzer}}

\newcommand{\nufnu}{$\nu f_{\nu}$}

\slugcomment{Paper status: ApJ in press}

\shorttitle{Optically--Dull AGN}
\shortauthors{Rigby \etal}

\begin{document}

\title{Why X-ray--Selected AGN Appear Optically Dull}
\author{J.~R.~Rigby\altaffilmark{1}, G.~H.~Rieke\altaffilmark{1}, 
J.~L.~Donley\altaffilmark{1}, 
A. Alonso-Herrero\altaffilmark{1,2}, \&
P.~G.~P\'erez-Gonz\'alez\altaffilmark{1}
}

\altaffiltext{1}{Steward Observatory, University of Arizona, 933 N. 
Cherry Ave., Tucson, AZ 85721}
\altaffiltext{2}{Instituto de Estructura de la Materia, Consejo Superior
de Investigaciones Cient\'{i}ficas, E-28006 Madrid, Spain} 

\email{jrigby@as.arizona.edu}

\begin{abstract}
We investigate why half of X-ray--selected active galactic nuclei 
(AGN) in deep surveys lack signs of accretion in their optical spectra.  
The majority of these ``optically--dull'' AGN are no more than $\sim6$ times 
fainter than their host galaxies in rest-frame R-band; 
as such, AGN lines are unlikely to be
overwhelmed by stellar continuum in at least half the sample.
We find that optically--dull AGN have the mid--infrared
emission and L$_x$/L$_{IR}$ ratios characteristic of local Seyferts, 
suggesting that the cause of optical dullness is not missing UV--optical continua.
We compare the morphologies of 22 optically--dull and 9
optically--active AGN at $0.5<z<0.8$, and find that optically--dull
AGN show a wide range of axis ratio, but optically--active AGN have
only very round axis ratios.  We conclude that hard X-rays select AGN
in host galaxies with a wide range of inclination angle, but only 
those AGN in the most face-on or spheroidal host galaxies show optical emission lines.  
Thus, extranuclear dust
in the host galaxy plays an important role in hiding the emission lines of 
optically--dull AGN.

[This manuscript uses low--resolution figures.  A full--resolution
manuscript can be obtained at
\verb8http://satchmo.as.arizona.edu/~jrigby/optdull.ps8] 
\end{abstract}


\section{Introduction}
\label{sect:intro} 

Deep X-ray surveys have found large numbers of active galactic nuclei
(AGN) at $z\sim1$, as predicted by models of the X-ray
background (e.g. \citealt{settiwoltjer,comastri95}).
  Most of these AGN have hard X-ray flux ratios that
indicate obscured accretion, which is consistent with the hardness of
the X-ray background.  In the local universe, AGN that are similarly
X-ray obscured have optical high-excitation emission lines that
classify them as Seyfert 2 galaxies.
Surprisingly, in deep surveys, $40$--$60\%$ of X-ray--selected AGN show  
{\bf no evidence} of nuclear accretion in optical spectra 
(see discussion and references in \citealt{moran}.)
Such sources are variously termed  ``optically--dull'', 
``optically--normal'', or ``X-ray--Bright Optically-Normal (XBONG)''.
By this, one means that these galaxies 
\emph{lack evidence for accretion activity} in optical spectroscopy.
They are therefore quite unlike a ``normal'' AGN, i.e. a QSO or Seyfert 
galaxy.\footnote{Terminology for deep X-ray sources can be confusing.
``Optically dull'' or ``optically normal'' AGN are X-ray--selected AGN that
lack AGN emission lines, but otherwise have typical host galaxies.
They are an entirely different population than the optically--bright
X-ray--faint sources in deep surveys (in which the X-rays come from 
X-ray binaries, not AGN) which are sometimes called ``normal galaxies.''}

Thus, what is so interesting about these AGN is precisely that 
they are ``dull'':  How can nuclei produce high X-ray luminosities 
(most have $41 < \log L_x < 44$~\ergss\ in rest-frame 2--8~keV)
that clearly require power from nuclear accretion, 
yet not show optical evidence of accretion?
The puzzle deepens when we consider that optically-dull AGN are rare
in the local universe.  As we discuss in \S\ref{sect:lowzhighz},
at most 10--$20\%$ of local hard X-ray--selected AGN are optically
dull.  Has the AGN population evolved?

In this paper, we test three explanations for optical dullness:
a) these AGN have weak ionizing continua, which do not excite the narrow
   line regions (e.g.~\citealt{barger2001}.)  
b) these are faint AGN in bright galaxies that overwhelm the AGN lines;
c) the host galaxies of these AGN have obscured their AGN lines.  
Explanations b) and c) are effects 
seen in low--redshift samples, as we summarize in \S~\ref{sect:hypotheses}.  
Explanation a) would require the AGN
population to evolve strongly, from optical--dullness at $z\sim1$ to optical
activity at $z\sim0$.  
Before appealing to AGN evolution,
we should consider whether explanations b) and c),
which are motivated by the behavior of local Seyferts, can fully explain
optical dullness.

\section{Insight from low--redshift Seyfert samples}
\label{sect:hypotheses}

Samples of nearby Seyferts suggest two likely causes for optical dullness in AGN.
The first possible
explanation is that optically--dull AGN 
have been observed in ways that dilute or hide their
optical activity.  This explanation was advanced by \citet{moran},
who observed 18 local Seyfert~2s with large apertures (to simulate
observations at high redshift), and demonstrated that in 11 cases, the
nuclear activity was drowned out by stellar light.  (Nine of those
sources showed early--type spectra, and two showed starburst spectra). 
This ``dilution'' hypothesis predicts that optically--dull AGN should
inhabit high--luminosity host galaxies, which generate enough
continuum to drown out the AGN lines.  

Another possible explanation is that the narrow--line regions of
optically--dull AGN have been obscured by their host galaxies.  This
hypothesis is motivated by selection effects observed in local Seyfert
samples.
\citet{keel} first demonstrated a deficiency of nearby Seyfert 1s in 
edge-on host galaxies.  
\citet{mcr95} later showed that much larger samples of nearby 
optically--selected Seyfert 1 and 2 AGN are biased against having 
inclined spiral hosts. They found samples of soft X-ray--selected 
QSOs to be  similarly biased.  By contrast,
they found that nearby samples of Seyferts selected by hard 
X-rays or 12~\micron\
emission showed relatively flat distributions of axis ratio from 0.2 to 1,
which is the expectation for randomly-oriented disk galaxies.
Based on the known spatial scales of narrow-line regions, and on the
differing biases of soft and hard--X-ray selected samples, they
concluded that many Seyfert galaxies have several magnitudes of
A$_V$ at scale-heights of $\ge100$~pc---which puts the
extinction outside a classical parsec-scale ``torus'', and rather
in a ``circumnuclear'' region at larger distances.  
They proposed that molecular rings of star-forming disturbed gas 
could account for the obscuring column. 
\citealt{almu03} also invoked A$_V \la 5$ of galactic extinction
outside the torus, to explain the nuclear SEDs of optically--classified
Seyfert 1.8 and 1.9 AGN.  In addition, \citep{malkan} found that
Seyfert~2 nuclei show more dust absorption than Seyfert~1
nuclei, which they proposed may be due to dust from irregular 
structures at $>100$~pc scales.
Thus, material in the circumnuclear region (100~pc scales) and at 
larger distances in the host galaxy can provide appreciable absorbing
column for the nuclear source.
This low redshift work suggests that at $z\sim1$, where Chandra
observes rest-frame hard X-rays, the deep surveys
should select AGN independent of inclination.  Inclined disk galaxies 
within these surveys should have preferentially weak
broad- and narrow-lines. 
The relatively large fraction of optically--dull
AGN at high redshift then arises from the comparatively low ratio of 
signal to noise in spectra of those faint objects, combined with the 
poor physical resolution to isolate their nuclei. 
  Are these trends observed?

\section{Sample selection and data}
\label{sect:data}

To investigate optically--dull behavior, we need a uniform sample of
such AGN at $z\sim1$.  
We start with the \citet{szokoly} catalog of 
X-ray--selected AGN with spectroscopic redshifts, all in the Chandra
Deep Field South (CDFS).  These AGN were originally selected from the
Chandra 1 Ms observations \citep{giacconi, alex}, and were
followed up with optical spectroscopy \citep{szokoly} obtained with the FORS1/FORS2
at the VLT ($R=5.5$~\AA\ pix$^{-1}$).  In this paper we use both the 
\citet{szokoly} catalog and electronic versions of their published spectra, 
available online\footnote{http://www.mpe.mpg.de/CDFS/data/}.

From the spectroscopic catalog, we select AGN with X-ray luminosities
that indicate AGN activity (which corresponds to Szokoly X-ray
classifications of AGN-1, AGN-2, QSO-1, or QSO-2).  We then select the
subset of those sources that have non--active optical spectral
classifications: either ``LEX'' (sources with low--excitation narrow
emission lines, as expected from star formation) or ``ABS'' (sources
with stellar absorption--line spectra).
These criteria select 45 sources, all of which have a spectroscopic 
redshift; 34 of these redshifts are listed as ``reliable'' (flag Q$=3$).
Nine of the 45 sources have ``ABS'' spectra, 
and the rest have ``LEX'' spectra.
Table~\ref{tab:sample} lists source positions, redshifts, 
optical spectral classifications, and X-ray classifications, all from 
\citet{szokoly}.  We use source identification numbers from \citet{alex}, 
abbreviated ``AID''.

We also list, in table~\ref{tab:sample}, 
several derived quantities for each source:
1) rest--frame 2--8 keV luminosity
(using the photon index inferred from the observed
0.5--2 keV / 2--8 keV flux ratio, and normalizing by the observed 
2--8 keV flux);
2) the absorption-corrected rest-frame 
2--8 keV X-ray luminosity, which we find by extrapolating the 
observed 2--8 keV flux assuming an intrinsic photon index $\Gamma=2$,
where $f_{\nu} \propto \nu^{1 - \Gamma}$;
and 3) the estimated column density N$_H$, which would 
make an intrinsic $\Gamma=2$ power-law spectrum at the source's redshift 
show the observed 0.5--2 keV / 2--8 keV flux ratio.
When the source was not detected in the hard (2--8 keV) band, we
quote upper limits on $N_H$, L$_x$, and absorption--corrected L$_x$.
When the source was not detected in the soft (0.5--2 keV) band, 
we quote an upper limit on  $N_H$.
The median $N_H$ errorbars, from  the uncertainty in the X-ray flux 
ratio, are $-29\%$ and $+39\%$.    

From the same catalog, 
we also create a comparison sample of ``optically--active AGN'':
X-ray--selected AGN that have either broad emission lines (``BLAGN'')
or high-excitation narrow emission lines (``HEX'') according to
\citet{szokoly}.  Observed and derived quantities are listed
in table~\ref{tab:sample-optact}.  The optically--dull and 
optically--active
samples have comparable R-band magnitude distributions, though on average,
the optically--active sources are somewhat fainter.
Multi--object spectroscopy was used to obtain the spectra, and 
exposure times were adjusted based on the R-band magnitude of the source.
Therefore, the optically--dull and optically--active samples have 
spectra with comparable distributions of signal--to--noise.

Having selected a sample, we assembled a multiwavelength 
photometric database.  Using \spitzer\ \citep{spitzer}, we obtained 
IRAC \citep{irac} images of the CDFS with $500$~s of integration.   
The images were  
reduced by the {\it Spitzer} Science Center using the standard pipeline.
We also obtained MIPS \citep{mips}
$24$~\micron\ scan map images with a total integration time of $\sim 1400$~s
per position, nominally composed of $120$ individual sightings per source. 
These data were reduced using the instrument team data analysis tool 
\citep{gordon}, creating the image presented by \citet{rigbyx24}. 

We add the following optical and near--infrared imagery:
ACS/HST \emph{bviz} images from GOODS \citep{goods_acs}; 
\emph{RIz} frames from the Las Campanas Infrared Survey \citep{marzke};
\emph{BVRI} images released by the ESO Imaging Survey \citep{arnouts};
\emph{JK} images from GOODS \citep{goods_acs}; and 
\emph{JK} images from the EIS Deep Infrared 
Survey\footnote{http://www.eso.org/science/eis/surveys/strategy\_EIS-deep\_infrared\_deep.html}.

Unlike optically--faint AGN \citep{rigby_optfaint}, optically--dull
AGN have relatively bright, unambiguous optical and near--infrared
counterparts.  Therefore, creating SEDs is straightforward.  The
result is closely--sampled, deep photometry from $0.4$ to $8$~\micron,
with additional coverage at $24$~\micron.  We classify the
$0.4$--$8$~\micron\ SEDs into 5 categories, based on the 
\citet{devriendt} templates: old stellar population;
young stellar population; intermediate-age stellar population;
flat in $\nu f_{\nu}$; or rising in $\nu f_{\nu}$ with increasing 
wavelength.  We deliberately
ignore $24$~\micron\ flux density when classifying, as it can be elevated by
star formation or accretion.  Table~\ref{tab:sample} lists the SED
classifications.

Using these data, we now examine the three most likely explanations for
optical dullness:  a) weak ionizing continua; b) dilution by the host
galaxy continuum; and c) obscuration.

We adopt an $\Omega_{m}=0.26$, $\Omega_{\Lambda}=0.74$, $h_o = 0.72$ 
cosmology throughout.  When we take luminosities and absolute 
magnitudes from the literature, we convert them to this cosmology.

%

\section{Do optically--dull AGN have weak ionizing continua?}
\label{sect:weak-continuua}

One of the proposed causes for optical dullness is intrinsically weak 
ionizing continua, such that the narrow line regions are not excited.  
We perform two tests of this hypothesis.

\subsection{Do the low-obscuration AGN have big blue bumps?}

First, do low--X-ray obscuration AGN show the ``big blue bumps''
\citep{filippenko} attributed to normal, UV--bright accretion disks?  
Such low--X-ray obscuration AGN are
likely to have low--extinction lines of sight to the nucleus, and thus
are likely to show bright blue continua if they have
accretion disks.  However, 
differential extinction along the different lines of sight
to the optical and X-ray components could result in a lack of big blue bumps.
How many of the X-ray--soft (and therefore presumably low
obscuration) AGN in the sample have indications of ``big blue
bumps''?
Of the 10 AGN with lowest estimated column densities 
($N_H < 3\times 10^{22}$~\cmsq),
only 2 have flat \nufnu\ SEDs;   
the rest have stellar SEDs that fall quickly in the blue.\footnote{Nine of the ten
sources have reliable redshifts according to \citet{szokoly}.}  So among
these presumably low-obscuration AGN, big blue bumps appear to be rare. 
This is consistent with dust extincting
the blue continuum, or with an intrinsically weak blue continuum, although
it is more difficult to explain under the dilution hypothesis.

There are indeed  14 sources with flat or probably flat SEDs---if
these are not the lowest--column sources, what are they?
Of the 14, 7 have low--reliability
spectral quality flags \citep{szokoly}; emission lines may well show up in
better--quality spectra, and thus these sources may not be true 
``optically--dull AGN''.

Thus, blue non--stellar SEDs are not common in the optically--dull sample, 
consistent with their having weak UV continua but also with their having UV
of normal strength but obscured along our sightline.  We need another test to
distinguish these possibilities.

\subsection{Do the SEDs of optically--dull AGN show normal mid-IR emission?}
\label{sect:seds}

If the UV continua of the optically--dull objects really are weak,
then the mid--IR emission should be abnormally weak for AGN.
Local Seyferts characteristically have bright mid-IR emission
(c.f. \citealt{spinmalk89}, \citealt{extended12um}, \citealt{bolo12um},
\citealt{maiolino95}) that is attributed to heating of dust by UV
energy from the central engine.
Assuming similarity in other regards, AGN with weak UV should also have
weak IR.  However, since dust heated by star formation also emits in the 
mid--infrared, we must choose our tests and samples carefully.

To first order, most of the rest--frame optical/near--IR SEDs 
in the sample are stellar.
We select two subsamples that are unlikely to be dominated in
the mid--infrared by emission from star-formation.  As subsample A, we
select the 9 optically--dull AGN whose optical spectra are dominated
by absorption lines.  Of these, seven have old stellar 
SEDs.\footnote{The other two, AID 126 and 247, have power-law SEDs,
but both have low--quality spectra (\citet{szokoly} flags of 1.0 and
0.5, respectively), and thus may not have been classified as ABS given
better spectra.}  This subsample should be the least contaminated by
star formation.
We create subsample B by selecting optically--dull AGN whose SEDs (in
observed $0.4<\lambda<8$~\micron) are dominated by old stars,
but whose spectra show low--excitation emission lines that indicate
some HII regions are present.  Eight optically--dull sources match
these criteria.

\begin{figure}
\figurenum{1}
\includegraphics[width=2.5in]{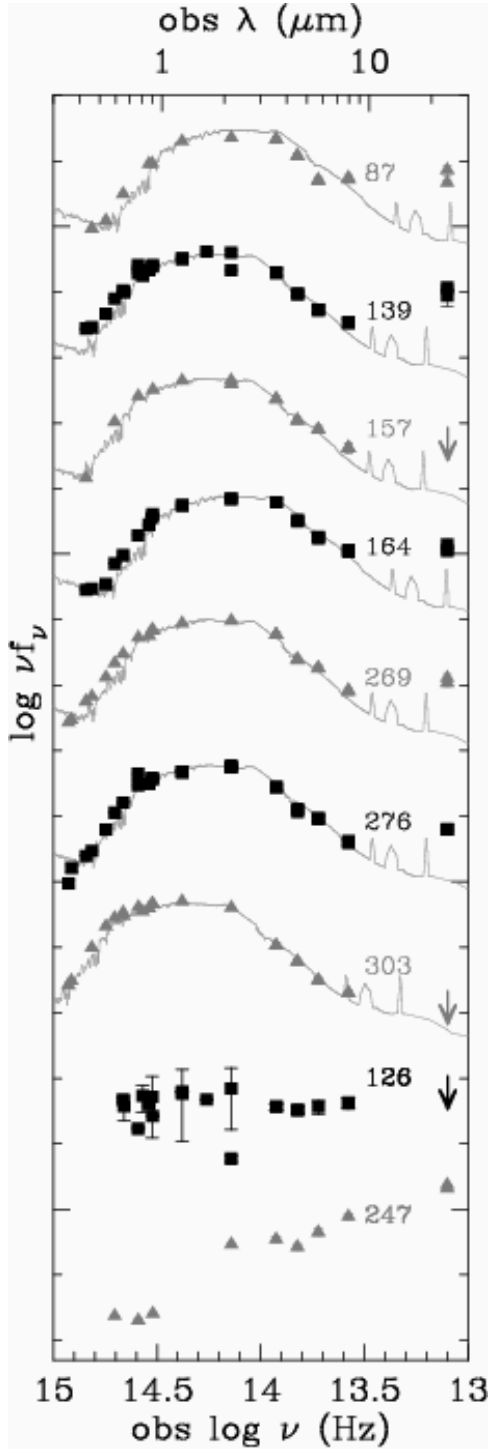}
\figcaption{Spectral energy distributions of subsample A 
(optically--dull AGN with absorption-line (ABS) spectral 
classifications.)  Wavelengths are as observed.
For comparison, we plot templates of the SA galaxy VCC~1003 in the Virgo cluster
\citep{devriendt}.}
\label{fig:optnorm_abs_seds}
\end{figure}

\begin{figure}
\figurenum{2}
\includegraphics[width=2.5in]{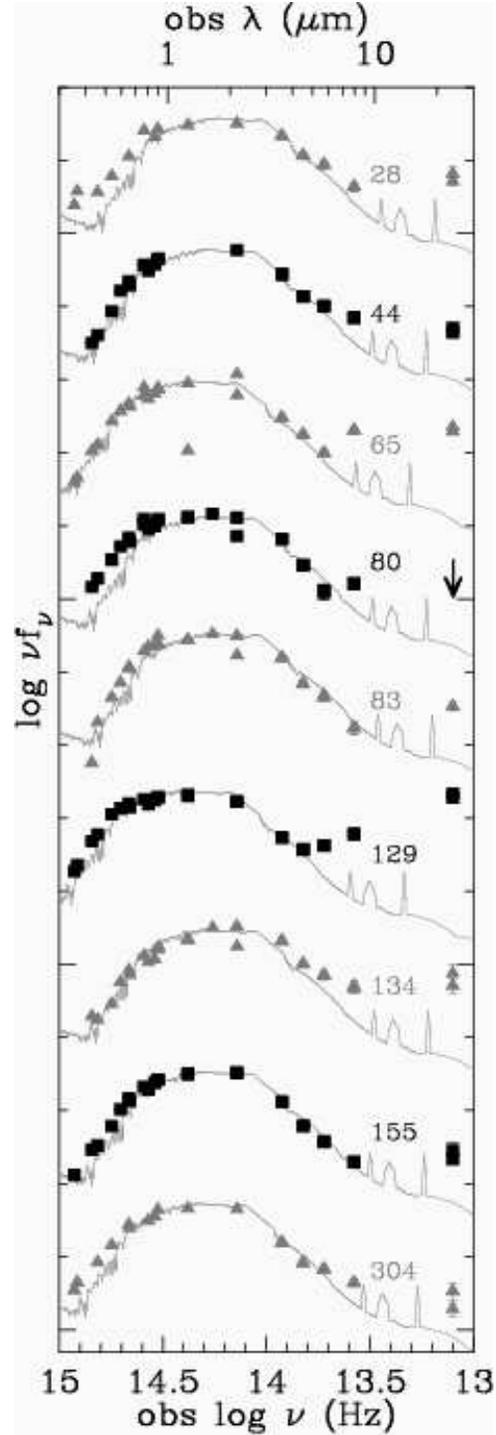}
\figcaption{SEDs of subsample B.  Wavelengths are as observed.
For comparison, we plot the SA galaxy VCC~1003 in the Virgo cluster 
\citep{devriendt}. }
\label{fig:optnorm_lex_seds}
\end{figure}

\begin{figure*}
\figurenum{3}
\vspace{-0.377in}
\includegraphics[width=4.4in]{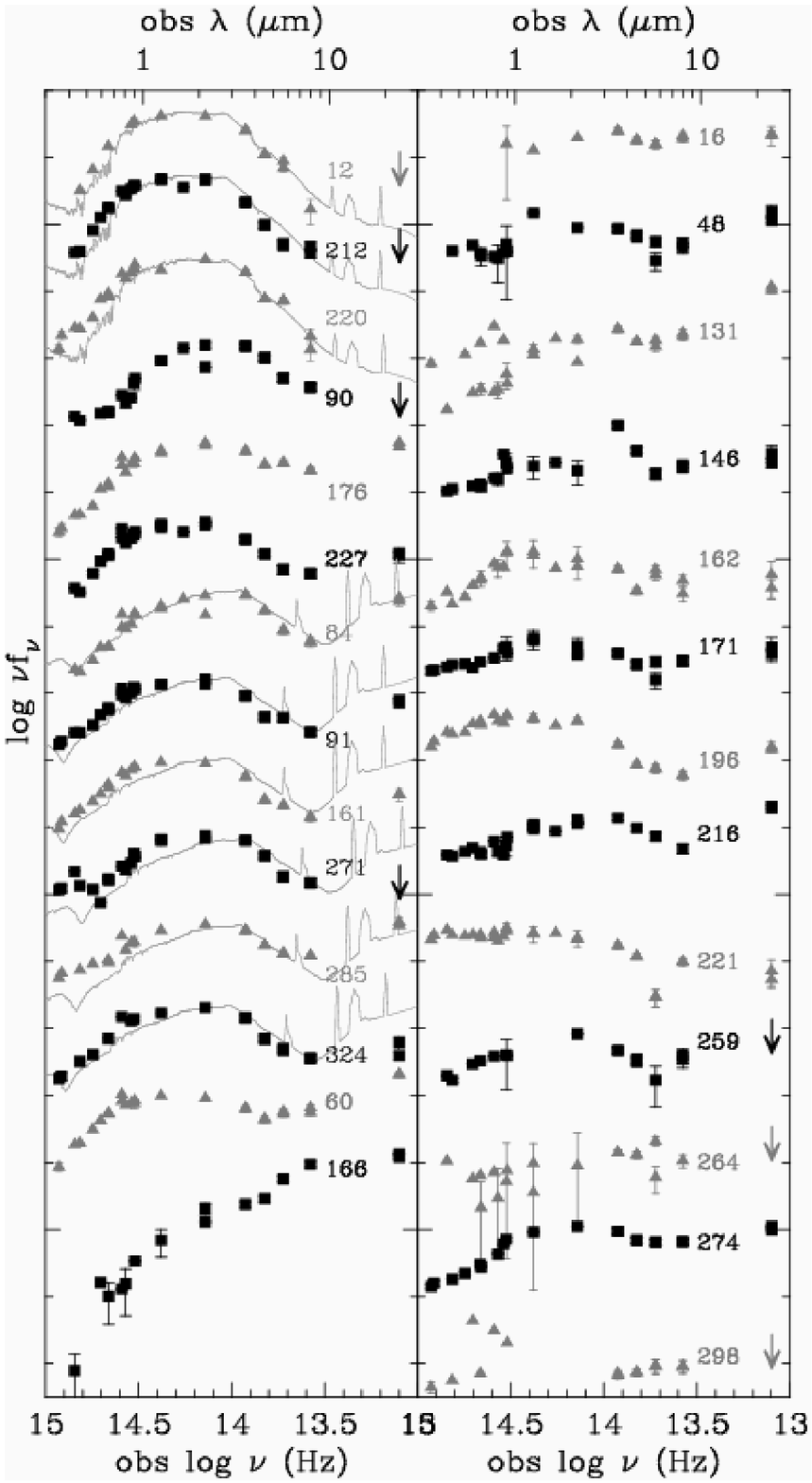}
\figcaption{SEDs of the remaining optically-dull AGN.  Wavelengths 
are as observed. Left panel, top to bottom:  3 old stellar pop SEDs; 
3 intermediate SEDs; 6 young stellar pop SEDs;
and 2 non-stellar rising-spectrum SEDs.  
Right panel:  the non--stellar flat SEDs.
For comparison, we plot the SA galaxy VCC~1003 with the old--type SEDs,
and M82 with the young--type SEDs \citep{devriendt}. }
\label{fig:optnorm_other_seds}
\end{figure*}

In figures~\ref{fig:optnorm_abs_seds}, \ref{fig:optnorm_lex_seds}, and
\ref{fig:optnorm_other_seds} we plot the SEDs of the optically--dull
AGN.  Compared to an Sa galaxy template \citep{devriendt}, 
5 of 7 subsample A (ABS) sources show elevated $24$~\micron\ emission, 
as do 6 of 8 subsample B sources.  Thus, this test supports the 
presence of significant UV luminosity.
However, this simple test is not definitive because the 
templates do not reflect the observed range of SEDs observed with 
\spitzer.  Additionally, the SED test is not clear-cut for the 
subsample B sources, which by definition contain some low--level star 
formation that could elevate the $24$~\micron\ flux density.

\subsection{Do optically--dull AGN have stellar or AGN-like H$\beta$/24~\micron\ ratios?}
\label{sect:Hbeta}
Since Balmer emission and mid--infrared flux density are both strongly
influenced by the star formation rate, we can use the observed
H$\beta$ emission--line flux to predict the observed $24$~\micron\
flux density predicted by star formation, and test whether the 
optically--dull AGN violate this expectation (by having 24~\micron\ excess.)

We obtain optical emission line fluxes using the VLT spectra of
\citet{szokoly} discussed in \S~\ref{sect:data}.  For the two
subsamples, the VLT spectra cover [\OIIlong], and generally cover all of
the Balmer series except H$\alpha$. 
In subsample A, no source shows Balmer emission; 4/7 show Balmer 
absorption, and 3/7 show non-existent or very weak Balmer lines.
Only 1 of 8 subsample B sources show H$\beta$ emission; 
the rest have non--detected Balmer lines.  

We fit continua at the Balmer lines and sum line fluxes.  When Balmer
lines are not detected or appear in absorption, we take a conservative
upper limit on emission flux as f(H$\beta$)=$3\times10^{-16}$\ergss
(which is the five$\sigma$ limiting H$\beta$ emission flux of the spectrum 
with lowest signal-to-noise ratio).
We deredden the Balmer fluxes by a standard extinction 
value\footnote{This corresponds to A$_{H\alpha}=1.1$ and
A$_{H\beta}=1.57$.  It is the extinction value applied by
\cite{roussel}, and is the average extinction observed by
\citet{rob98}.} of E(B-V)$=0.435$.

\begin{figure}
\figurenum{4}
\includegraphics[angle=0, width=3.4in]{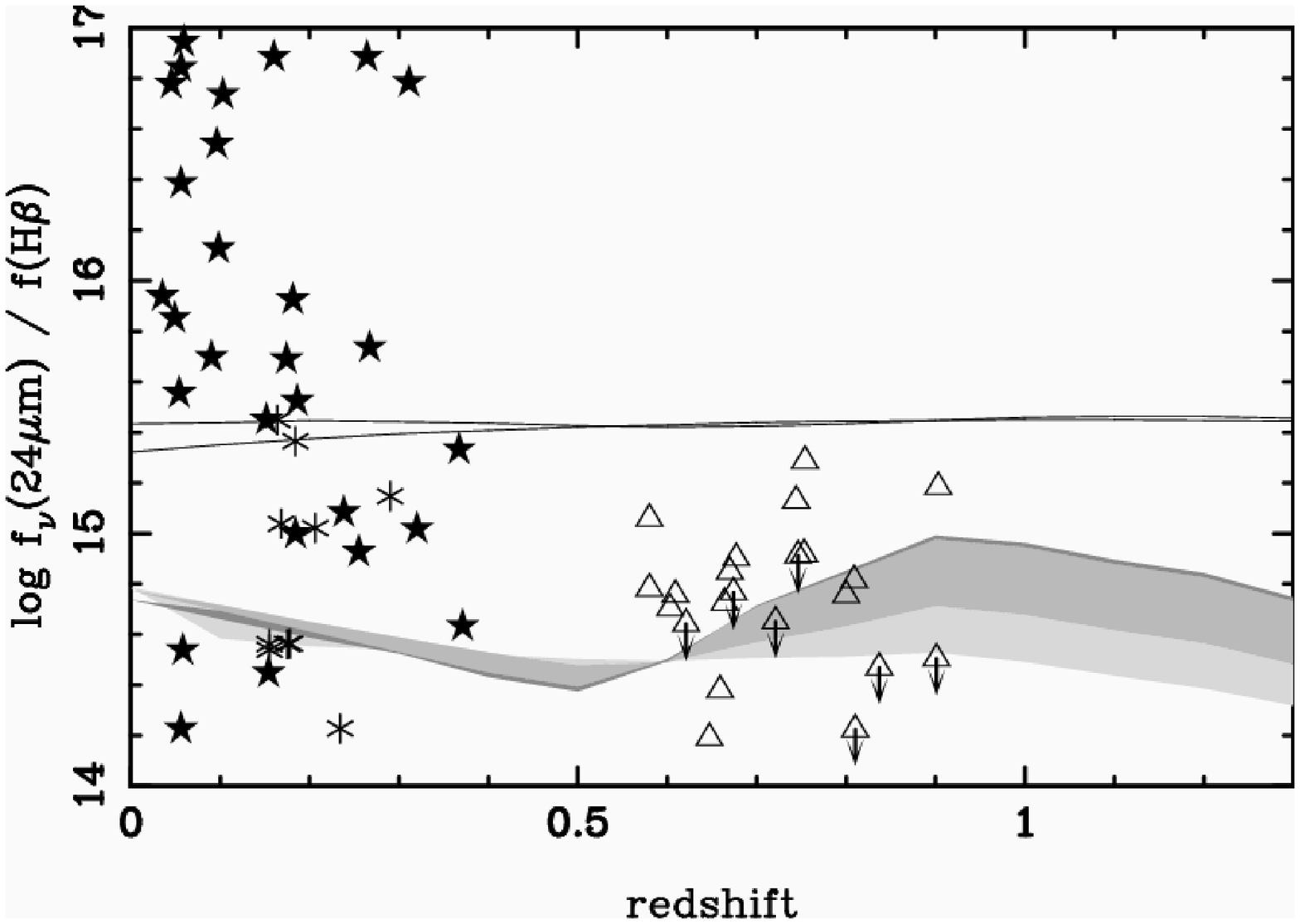}

\includegraphics[angle=270, width=3.4in]{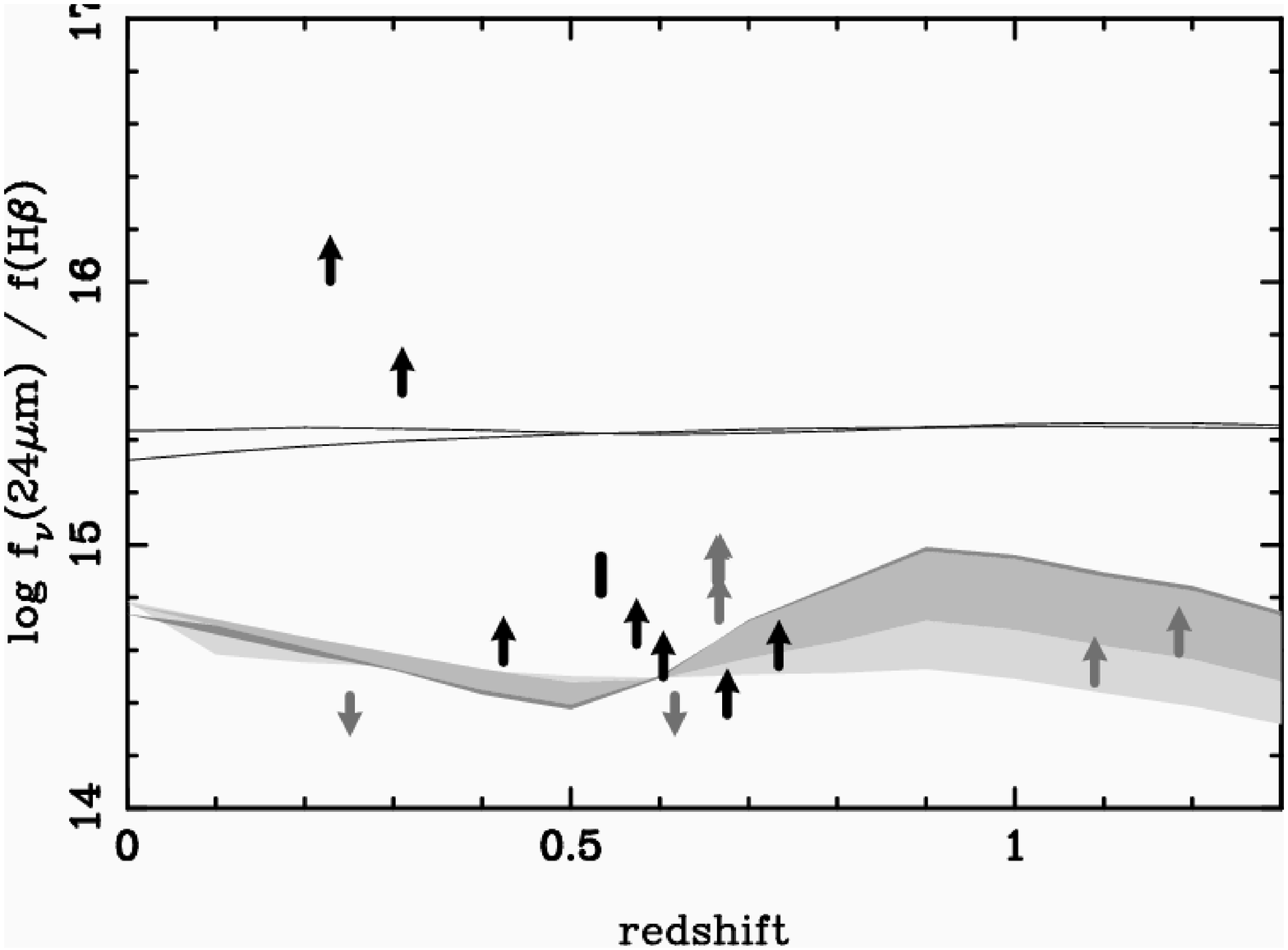}
\figcaption{The observed $f_{\nu}(24~\micron)/f(H\beta)$ ratio. 
The upper panel shows the comparison samples:  PG quasars 
(\emph{asterisks}, 24~\micron\ data from D. Hines, private communication, 
H$\beta$ from the literature);
radio quasars and FRII radio galaxies (\emph{filled stars}, 24~\micron\ data
from \citealt{yong}, H$\beta$ from the literature);
star-forming galaxies from the Extended Groth Strip 
(\emph{triangles}, unpublished 24~\micron\ data, H$\beta$ from \citealt{lilly03});
and star--forming galaxy models with log~L$_{bol}=$9--10~\Lsun 
(\emph{darkest shaded region});  
log~L$_{bol}=10$--$11$~\Lsun\ (\emph{lighter shaded region});
and log~L$_{bol}=$11--12~\Lsun\ models (\emph{lightest shaded region}),
all from \citet{charyelbaz}.
Solid lines show the expected K-corrections for median PG QSOs,
from \citet{elvis} and \citet{haas}, with arbitrary normalization.
The bottom panel shows the optically--dull AGN from subsample A
(\emph{grey arrows}) and subsample B (\emph{black arrows and bar}).
The units of the y-axis are mJy$/$(\cgsflux). 
}
\label{fig:hbeta}
\end{figure}

In figure~\ref{fig:hbeta} we plot the observed
$f_{\nu}(24~\micron)/f(H\beta)$ ratios for subsamples A and B (lower
panel), as well as for several comparison samples (upper panel).  The
first comparison sample is a sample of star-forming galaxies at
$z\sim0.7$, generated by cross-correlating the H$\beta$ catalog of
\citet{lilly03} with a $24$~\micron\ catalog obtained from our GTO
\spitzer\ images of the Extended Groth Strip.  The second is a sample
of low-redshift PG quasars, with 24\micron\ flux densities taken from
the MIPS GTO AGN survey (D. Hines, private communication) and H$\beta$
fluxes from \citet{marziani}; these quasars are unlike our sample in
that they are generally unobscured AGN.  The third sample consists of
radio galaxies and quasars from \citet{yong}; the radio galaxies are
more like our optically--dull sample in that their nuclei are
obscured.

We also plot the \citet{charyelbaz} star-forming galaxy models with
bolometric luminosities from $10^9$~\Lsun\ to $10^{12}$~\Lsun\ (see
appendix for details on the K-correction), along with the K-correction paths
of two median PG quasar SEDs, the \citet{elvis} template and  the
normalized (at $\lambda_{rest} =$ 25~\micron), de-redshifted median photometry 
of the \citet{haas} sample.  

There are several points to take from figure~\ref{fig:hbeta}.

First, the galaxies from \citet{lilly03} are in the general range of the
predictions of the \citet{charyelbaz} models.  This
suggests that the empirical relation between H$\beta$ and
mid--infrared flux density in star-forming galaxies at $z\sim0$
\citep{roussel} holds reasonably well out to $z\sim1$.
 
Second, at least two optically--dull AGN have $f_{\nu}(24~\micron)/f(H\beta)$ 
ratios that are too high to be consistent with star formation.  These
are the highest of the upward--pointing arrows, i.e., those  with
$\log f_{\nu}(24~\micron)/f(H\beta) > 15.4$~mJy/(\cgsflux).
\emph{Thus, we see strong mid--infrared evidence of accretion in a few
optically--dull AGN} from the 24~\micron/H$\beta$ test.

Third, the $f_{\nu}(24~\micron)/f(H\beta)$ ratio \emph{is not} a
useful diagnostic of AGN versus star formation activity for most
of these sources.  The problem is, the range of ratios seen for
$0.5<z<1$ star-forming galaxies overlaps with the range seen in PG
quasars and radio galaxies.  Therefore, the ratio for the optically--dull
galaxies are consistent with AGN--powered IR excesses, but the
majority of sources have ratios that could also arise from star formation.


\subsection{Do optically--dull AGN have AGN-like Mid-IR to X-ray luminosities?}
\label{sect:lutz}
The previous two subsections have shown that some optically--dull
AGN appear to have AGN--powered excess 24~\micron\ emission.  
The rest are consistent with this possibility, but 
differential extinction and inadequate comparison templates make arguments based on
optical SEDs or H$\beta$ line strength inconclusive for many of these galaxies. 
Therefore, we now compare the mid--infrared and X-ray
luminosities of the optically--dull AGN, since both should be robust
to large amounts of obscuration.

\begin{figure}
\figurenum{5}
\begin{center}
\includegraphics[angle=270, width=3in]{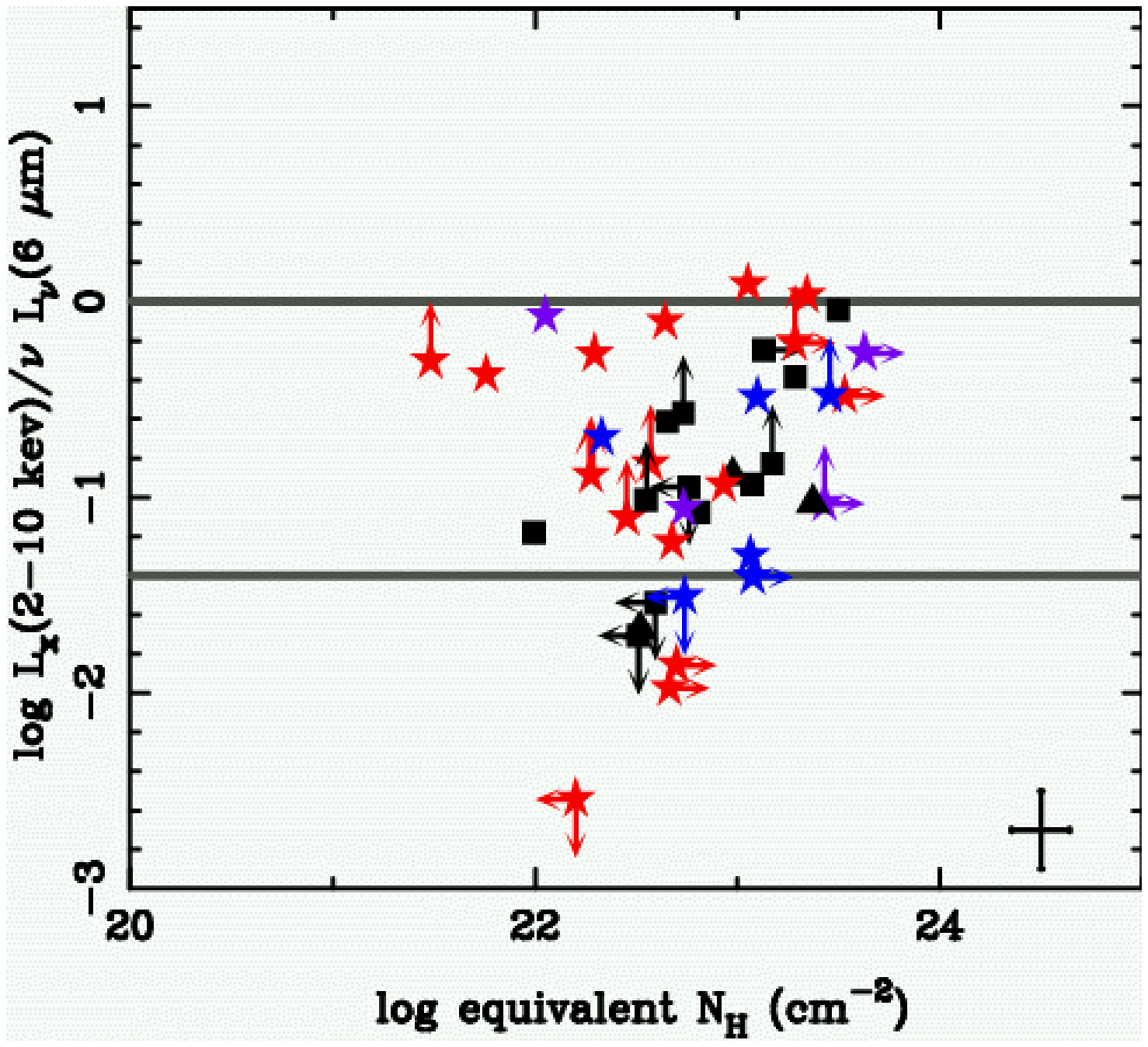}\\
\includegraphics[angle=270, width=3in]{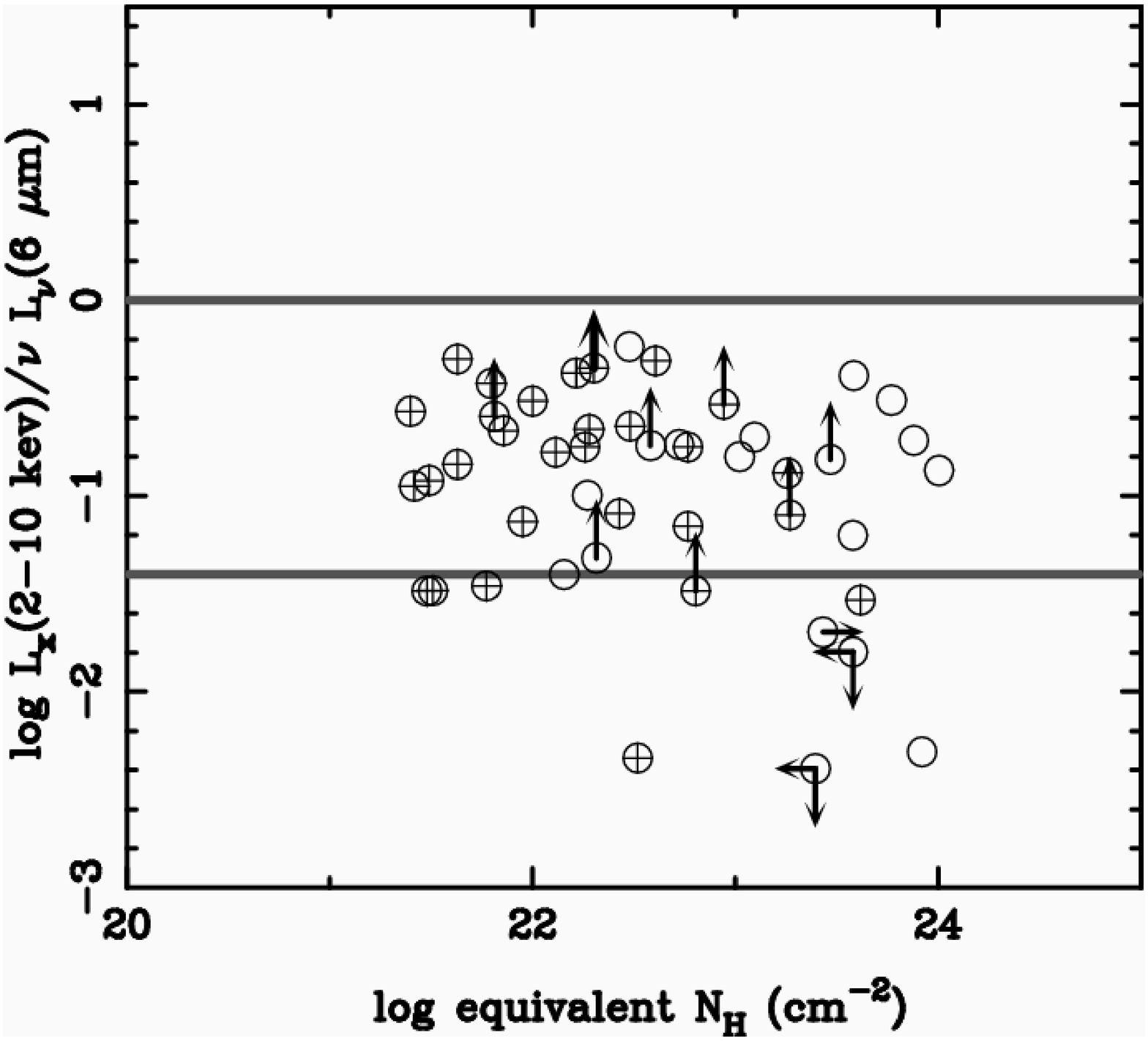}
\end{center}
\figcaption{Ratio of 6~\micron\ and X-ray luminosity versus inferred 
column density.
Optically--dull AGN are plotted in the upper panel, 
with symbols coded by SED type:
old stellar pop\textit{(red stars)}; 
intermediate \textit{(purple stars)};
young stellar pop\textit{(blue stars)}; 
flat in \nufnu \textit{(black squares)};
rising in \nufnu with increasing wavelength \textit{(black triangles)}.
Optically--active AGN, both 
BLAGN \textit{(crossed circles)} and narrow-line AGN \textit{(open circles)} 
are plotted in the lower panel.   Horizontal lines mark the approximate 
luminosity ratio range observed in the low--redshift sample of 
\citet{lutz} (see their figure~7.)  Star--forming galaxies without AGN
would have y-axis values in the range of -1.3 to -2.7, using data from 
\citet{zezas} for three galaxies and the 6\micron-to-FIR 
relation of \citet{elbaz2002}.
Representative errorbars are plotted.
}
\label{fig:lutz}
\end{figure}

In figure~\ref{fig:lutz} we plot the ratio of the
absorption--corrected rest--frame 2--10 keV luminosity to the
rest-frame 6~\micron\ luminosity, as a function of estimated column density.  
We calculate rest-frame $\nu L_{\nu}$(6~\micron) by extrapolating from the 
observed 24~\micron\ flux density, assuming 
f$_{\nu} \propto \nu^{-1}$)\footnote{Very similar luminosities are obtained 
if we instead calculate 6~\micron\
luminosity by interpolating between the observed 24~\micron\ and
8~\micron\ flux densities.}.  
Figure~\ref{fig:lutz} should be compared with figure~7 of
\citet{lutz}, which is the corresponding plot for $\sim40$ nearby
active galaxies with \textit{ISO} spectra.

The first insight to gain from figure~\ref{fig:lutz} is that the
$z\sim 0.7$ AGN behave much like the low--redshift AGN of \citet{lutz}.
The range of $L_x/L_{IR}$ is similar (apart from a few outliers at low
$L_x/L_{IR}$).  \citet{lutz} showed that the luminosity ratio does not
depend on column density for local AGN; this finding constrains
unification geometry, since many unification models predict a
dependence on N$_H$.  The CDFS sample shows that $L_x/L_{IR}$ of AGN 
does not depend on column density at $z\sim0.7$, either, a result first
indicated by \citealt{rigbyx24} from flux ratios, and also demonstrated
by figure~9 of \citet{powerlaw}.)

The second insight is that the optically--dull and optically--active
AGN in the CDFS do not have significantly different $L_x/L_{IR}$ ratios.
Specifically, figure~\ref{fig:lutz} does \emph{not} show that
optically--dull AGN have abnormally low mid--infrared luminosities
given their X-ray luminosities.  This result, along with those of the
previous two section, argues strongly that \emph{optically--dull AGN
have Seyfert--like  mid--infrared emission.}

The third insight is that the sources with young and old stellar SEDs do not
have significantly different $L_x/L_{IR}$.  This tentatively suggests 
that the scatter in $L_x/L_{IR}$ for AGN is not
primarily caused by emission from star formation.  It
is unknown what causes the considerable scatter in $L_x/L_{IR}$; both
quantities should be reasonably robust to extinction and other
line-of-sight effects.  The observed scatter is larger than that
expected from photometric uncertainties alone (the median fractional
error in  $L_x/L_{IR}$ is $25\%$.)


Last, we note that the CDFS AGN include a handful with lower
$L_x/L_{IR}$ than seen by \citet{lutz}.  

Thus, we conclude that the optically--dull AGN have X-ray and mid-IR 
luminosities within the normal ranges seen for local Seyferts.  The SED and 
24~\micron/H$\beta$ tests are consistent with the behavior seen in the
$L_x/L_{IR}$ test, and in some cases independently corroborate it.  
We conclude that \emph{optically--dull AGN have the normal mid--IR emission
expected for Seyfert galaxies.}  This argues that optically--dull AGN have normal AGN
blue and UV continua,  to produce a normal AGN mid-IR luminosity.
Thus, the lack of big blue bumps in the X-ray--soft AGN is probably not caused
by intrinsically weak optical and UV continua. 


\section{How important is dilution to optical dullness?}
\label{sect:dilution}
We now estimate quantitatively the importance of ``dilution'' (bright 
galaxy continua overwhelming lines from faint AGN) in causing optical dullness.
Dilution has been argued to be important in local samples of weak--line AGN,
and so we first examine how similar these local AGN are to the $z\sim1$
optically--dull AGN.  Next, we examine the optical-to-X-ray flux and luminosity
ratios, to estimate the importance of dilution in our $z\sim1$ sample.

\subsection{Are optically--dull AGN at $z\sim1$ like weak-line AGN at $z\sim0$?}
\label{sect:lowzhighz}

Several studies have searched for local examples of optically--dull
AGN, since such sources are amenable to detailed follow-up.
These searches generally obtain optical spectroscopy for sources in
wide-area hard-band X-ray surveys, and identify the handful that lack
optical emission lines.  Several examples:
\begin{enumerate}
\item  The 2--10 keV--selected HEAO-1 sample \citep{piccinotti} 
contains 39 sources that are not galaxy clusters; 
30 of these have emission--line signatures of AGN, 
7 were not identified with known sources, 
and only 1 corresponds to an optically  non-active 
(i.e., an optically--dull) galaxy.
\item Follow-up spectroscopy of  5--10~keV Hellas Beppo-SAX sources 
found optical counterparts for 61/74 sources; 6 are apparently 
AGN without optical AGN signatures \citep{lafranca}
and have emission-line EW(H$\beta$)$\sim$ 1--5 \AA.    
\item The ASCA 2--10 keV Medium Sensitivity Survey identified 28 sources
at $z<0.2$ \citep{akiyama}.  All are broad-line, narrow-line, or BL Lac AGN, 
but five have weak emission lines (H$\beta$ and [\ion{O}{3}] equivalent widths 
below $10$\AA), and consequently might look optically-dull at 
higher redshift.
%
\item \citet{watanabe} followed up the hardest $2\%$ of the ASCA
detections, 
and found three ($z\sim0.05$) sources 
with N$_H\sim10^{23}$~\cmsq\ and abnormally low--luminosity Seyfert/LINER 
emission lines.  At redshifts typical of the deep surveys, these 3 sources
should look like optically--dull AGN.
\item Using a different search method, \citet{horn} correlated Chandra
images with spectroscopy from the Sloan Digital Sky Survey.  They
found no X-ray sources that lacked optical signs of nuclear accretion.
However, 4 of their 19 AGN would not be identified as AGN through
optical spectroscopy if their redshifts exceeded $\sim0.5$.
\end{enumerate}

Although it is not always clear for these studies whether additional
members of the sample would appear optically--dull at high redshift,
they agree that locally, optically--dull AGN are not common in samples
selected on the basis of X-ray emission.  

\begin{figure}
\figurenum{6}
\includegraphics[width=3.4in]{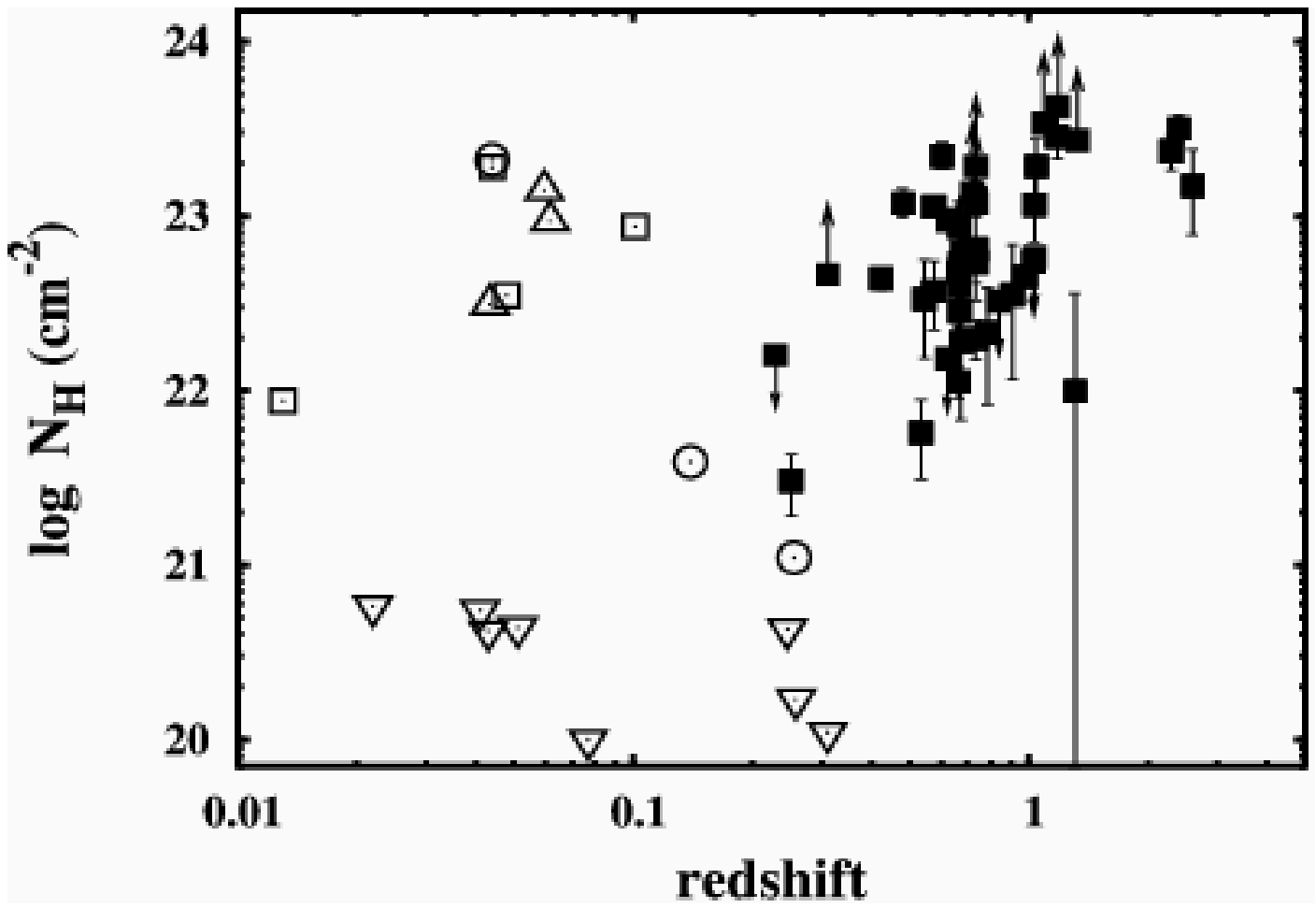}
\figcaption{Column densities of optically--dull AGN.  The low--redshift
samples are:  
XMM serendipitous sources from \citep{severgnini} \textit{(open circles)};
hard ASCA sources from \citet{watanabe} \textit{(up-pointing triangles)}; 
hard ASCA sources from \citet{akiyama} with emission line 
equivalent widths $<10$\AA\ \textit{(down-pointing triangles)}; and
5--10 keV BeppoSAX sources without AGN emission lines \citep{lafranca}
\textit{(open squares)},
where we have used the (1.3--4.5 keV)/(4.5-10 keV) count ratios 
derived from \citet{hellas} to estimate the column density.
The higher redshift sources are our CDFS sample \textit{(filled squares)}, with
N$_H$ values inferred from the flux ratio of the observed 2--8 keV and
0.5--2 keV bands (fluxes from \citealt{alex}).
}
\label{fig:NHz}
\end{figure}

In figure \ref{fig:NHz} we plot the column densities and redshifts of
these local AGN\footnote{\citet{horn} and \citet{piccinotti} do
not list flux ratios, so we cannot infer a column density.}. For
comparison, we also plot inferred column densities for the CDFS 
optically--dull AGN (see \S\ref{sect:data}).
Figure~\ref{fig:NHz} clearly demonstrates that many low--redshift
AGN with weak emission lines have very low ($< 10^{22}$~\cmsq)
column densities, much lower than the CDFS AGN.  Since they are
so much less obscured, they are not true analogues to the sources
in the deep fields, and so the importance of dilution to local weak--line
AGN should not argue for the importance of the effect in very different
AGN at $z\sim1$.

Thus, dilution may be important for the low--column,
local AGN with weak lines; but the much higher column densities of the 
CDFS AGN argue, at the very least, that these populations are not
similar.  This opens the possibility that extinction or other effects 
may be more important for the distant AGN than for local weak--line AGN.

\subsection{X-ray to optical flux and luminosity ratios}
\label{sect:lumrats}
\begin{figure}
\figurenum{7}
\includegraphics[angle=270, width=3.4in]{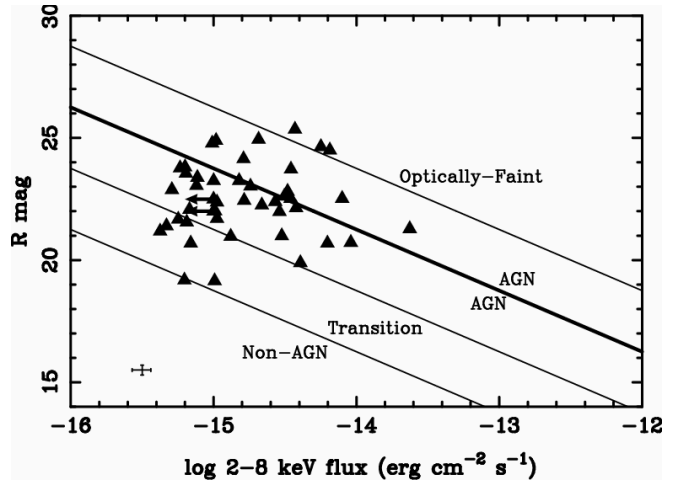}
\figcaption{Observed R-band magnitude and X-ray flux for the optically--dull AGN.  
Data are from  \citet{giacconi}.  Lines mark flux ratios of 
$F_x/F_R=$ 0.1, 1, and 10, following figure~3 of \citet{comastri}.
Representative errorbars are plotted.}
\label{fig:frfx}
\end{figure}

To further understand the importance of dilution to distant 
optically--dull AGN, in figure~\ref{fig:frfx}
we plot the observed R-band and 2--8 keV fluxes of the CDFS sample.
This figure should be compared to figure 3 of \citet{comastri}.
\citet{comastri} looked at ten $z\sim0.2$ optically--dull AGN from 
HELLAS2XMM and Chandra \citep{barger2001}, and found that most have
high $f_r/f_x$ ratios---which suggests that they have very luminous
host galaxies (which makes spectral dilution likely.)  In contrast,
our figure~\ref{fig:frfx} demonstrates that this effect is \emph{not}
common for the 45 CDFS optically--dull AGN: all but a handful have
optical/X-ray flux ratios that are typical of AGN.  Thus, this plot 
does not support the dilution hypothesis for optically--dull AGN in the CDFS.

\begin{figure}
\figurenum{8}
\includegraphics[width=9cm]{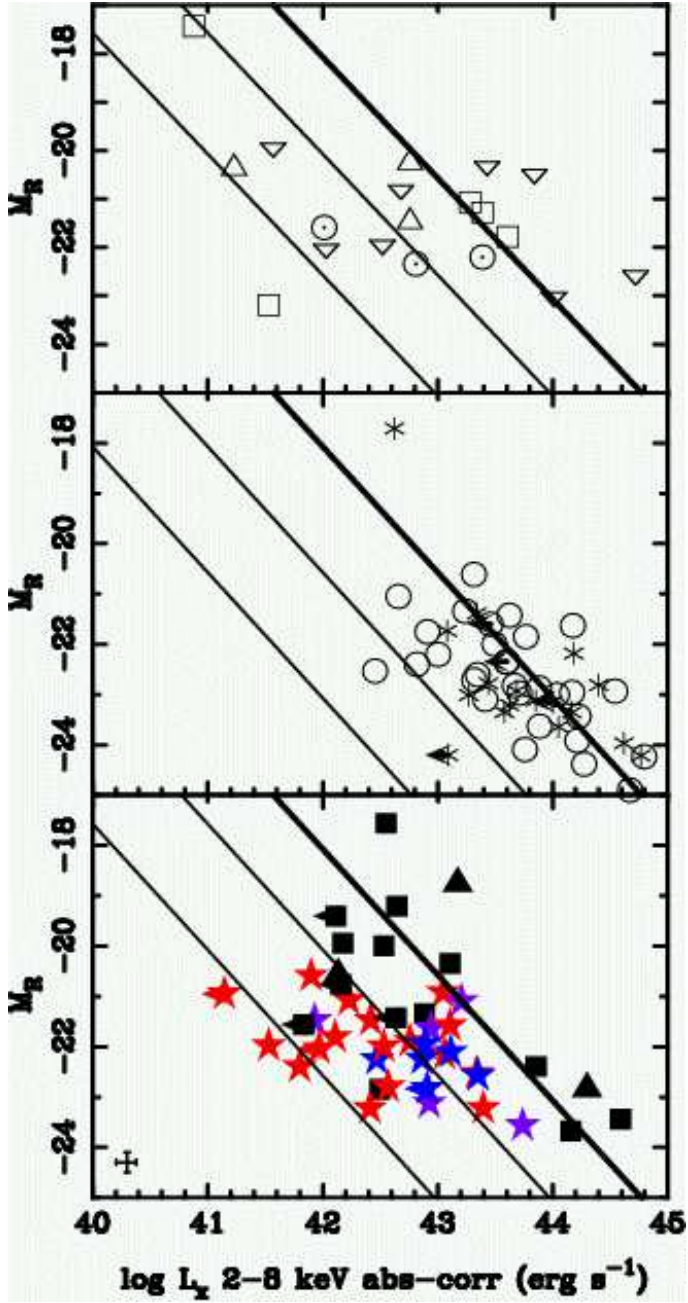}
\figcaption{Absolute magnitude M$_R$ versus absorption--corrected 
rest--frame 2--8 keV L$_x$. 
\textit{Top panel:}  Low--redshift weak-line AGN samples, with symbols
as in fig~\ref{fig:NHz}.  
\textit{Middle panel:}  Optically--active AGN from the CDFS,
both broad-line \textit{(asterisks)} and narrow-line \textit{(circles)}.
\textit{Bottom panel:}  Optically--dull AGN from the CDFS, symbols as in 
figure~\ref{fig:lutz}.
The plotted X-ray luminosity is absorption--corrected, in the rest--frame 
2--8 keV band, and is calculated from the observed 2--8 keV flux \citep{alex}
by  assuming the intrinsic spectrum is a $\Gamma=2$ power-law.  
M$_R$ is calculated by linearly interpolating between bands that bracket
the rest-frame R-band.  The thick diagonal line shows the luminosity ratio
of the composite \citet{elvis} QSO SED.  The thin diagonal lines mark
ratios with 10 and 100 times higher optical luminosity. 
Representative errorbars are plotted.
}
\label{fig:mr_lxabscor}
\end{figure}

However, figure~\ref{fig:frfx} is not an ideal diagnostic tool, since
it uses observed fluxes and thus does not K--correct.  Accordingly, in
figure~\ref{fig:mr_lxabscor} we plot the rest--frame absolute R magnitudes and
rest-frame  absorption--corrected 2--8 keV luminosities of the
optically--dull sample.  (We also plot the local--universe weak-line AGN from
figure \ref{fig:NHz}, and the optically--active AGN in the CDFS.)  
In each panel of figure~\ref{fig:mr_lxabscor}, 
the thick diagonal line shows the M$_R$/L$_x$
ratio of the composite radio-quiet QSO SED from \citet{elvis}.  The
thin diagonal lines mark flux ratios 10 and 100 times brighter in the
optical than the QSO $M_R/L_x$ ratio.  
Panel~b of figure~\ref{fig:mr_lxabscor} shows the M$_R$/L$_x$ ratios
of the optically--active AGN; $96\%$ of those sources lie above the
$10\times$ Elvis line; as such, their AGN should contribute at least $10\%$ of their
optical luminosities, and in many cases should dominate.
Comparison with panel~c then shows that 
$67\%$ of the optically--dull sources have M$_R$/L$_x$ ratios 
above the $10\times$ Elvis line.  
A more restrictive cut would be the $6.3\times$ Elvis line:
$94\%$ of the optically--active sources fall above it; and $56\%$
optically--dull sources do.

Thus, we conclude, based on figures~\ref{fig:frfx} and 
\ref{fig:mr_lxabscor}, that the R-band to X-ray luminosity ratios 
\emph{suggest that less than half} of the optically--dull AGN are dull because a
bright galaxy drowns out the AGN optical flux.  However such dilution
is likely at work in the optically--dull AGN whose 
total (AGN+host) optical luminosities are $\ga 6$ times brighter than 
the X-rays the AGN should be; this effect may be important for local
weak--line AGN.  Thus, we conclude that dilution is only a plausible
explanation for half or less of the CDFS optically--dull AGN.  
We need a cause for optical--dullness in the remaining half.



\section{Can Host Galaxy Obscuration Cause Optical--Dullness in AGN?}
\label{sect:morph}

Having shown that spectral dilution cannot account for the optical
dullness of half the sample, and that the ionizing continua are
unlikely to be missing given the normal mid-IR luminosities,
we now examine the role that host galaxy obscuration may play in 
causing optical dullness.
To do this, we examine the axis ratios of the optically--dull
and optically--active AGN in the CDFS.  
Most ($80\%$) of the optically--dull AGN lie within the ACS
GOODS survey described by \citet{goods_acs}.  
These deep, high--resolution
images provide morphology information even on distant sources.
In the following analysis, we use the i-band images from the 
v1.0 public release of the GOODS ACS images.
Measured axis ratios of most sources vary by only a few 
percent among the v, i, and z-band catalogs; therefore, 
we are free to choose the i-band, to optimize depth and spatial
resolution (sampling rest-frame optical light at $z\sim 0.7$).

\subsection{Measuring Axis Ratios}
\label{sect:morphba}
For each AGN, we measure the axis ratio, b/a, using two methods.

First, we simply take the b/a values from the GOODS v1.0 public
release i-band Source Extractor catalogs.  While Source Extractor
\citep{sextractor} is not optimized for morphological parameter
fitting, it is a widely-available tool that can quickly estimate axis
ratios for thousands of galaxies, and thus is of potential interest
for large surveys. 

Second, we iteratively fit multi-component Galfit models
(\citealt{peng}, Galfit v2.0.3b) to each source.  This method is
more time-consuming, but potentially more robust in its measurement of
b/a than Source Extractor.  We fit the following four Galfit models to
each source:
\begin{itemize}
\item a point source, i.e.~the HST PSF (which we created from stars in the 
   GOODS i-band images).
\item a single Sersic component.
\item a point source plus a Sersic component.
\item a deVaucouleurs component plus an exponential disk component.
\end{itemize}

We  adopted the b/a ratio of the best-fitting model, as follows:
\begin{itemize}
\item If the PSF provided the best fit, we considered the source 
to be unresolved, and so we have no knowledge of the true b/a ratio.
\item If the Sersic or PSF+Sersic models fit best, we adopted the
Sersic component's b/a.  
\item If the bulge+disk model fit best, and both components had effective radii
greater than the PSF FWHM, then we adopted
as b/a the weighted mean of the two component b/a values, weighted by
the fluxes of the model components.   
If only one component had $R_e >$ FWHM,
then we used the b/a value of that component.
(Since any $R_e <$ FWHM component is essentially unresolved, 
the b/a value galfit assigns it is not meaningful.)
(This weighting may over-estimate b/a in very inclined galaxies, since the
bulge will generally be rounder than the disk.)
\end{itemize}

When sources were fit well by more than one Galfit model, the axis
ratios of the models agreed well.  Also, the b/a values measured by
Source Extractor and Galfit are quite similar for each source, though
Galfit tends toward smaller b/a ratios since it can fit multiple
components.  We quote axis ratios from Galfit unless otherwise
indicated.

\subsection{Axis Ratios of the $0.5<z<0.8$ Subsample}
\label{sect:morph_subsample}
We can now compare the axis ratios of the optically--dull and
optically--active AGN.  We must take care because the luminosity and
redshift distributions of the two samples differ.  Specifically, the
optically--active AGN tend to be more luminous and lie at higher redshift,
and thus the HST images provide lower physical resolution; this means
that the optically--active sources are more likely to be unresolved or 
have limited morphology information.
 To minimize this bias, we consider
the subsample of optically--dull and optically--active AGN with GOODS
coverage that have redshifts between $0.5$ and $0.8$\footnote{All of the
optically--active sources in this redshift range have reliable redshifts,
and all but 3/22 of the optically--dull sources do.}. 
Because this
includes the ``redshift spike'' in the CDFS, the subsample contains more
sources than one would expect in an average field: there are 22
optically--dull AGN, and 9 optically--active AGN\footnote{Three are 
narrow-line AGN (88, 179, and 241), and the rest are broad-line AGN.}.
The absorption--corrected X-ray luminosity distributions of these 
two samples are comparable.

\begin{figure*}
\figurenum{9}
\vspace*{-0.01in}
\includegraphics[angle=270, width=6in]{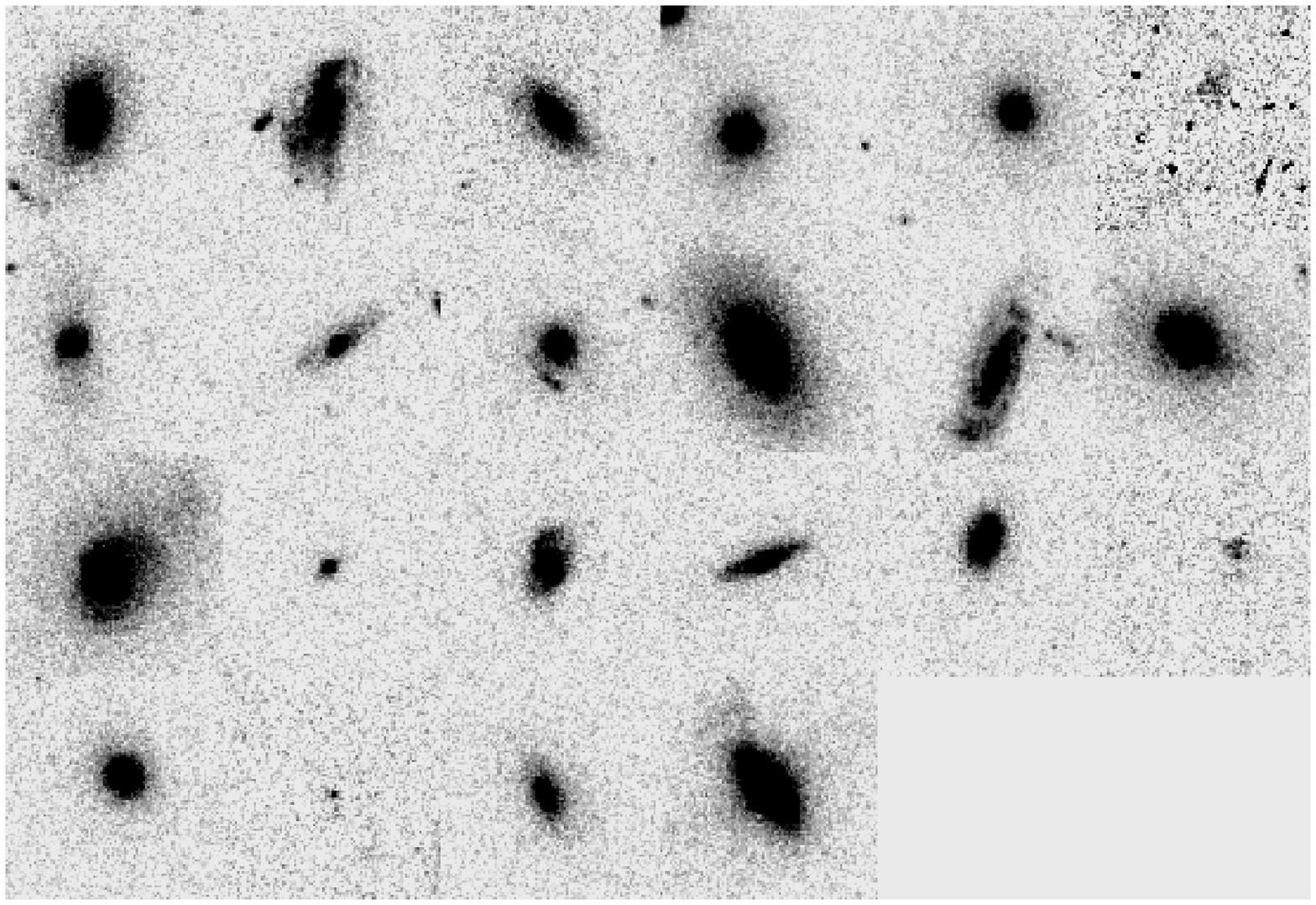}

\vspace*{0.05in}
\includegraphics[width=6in]{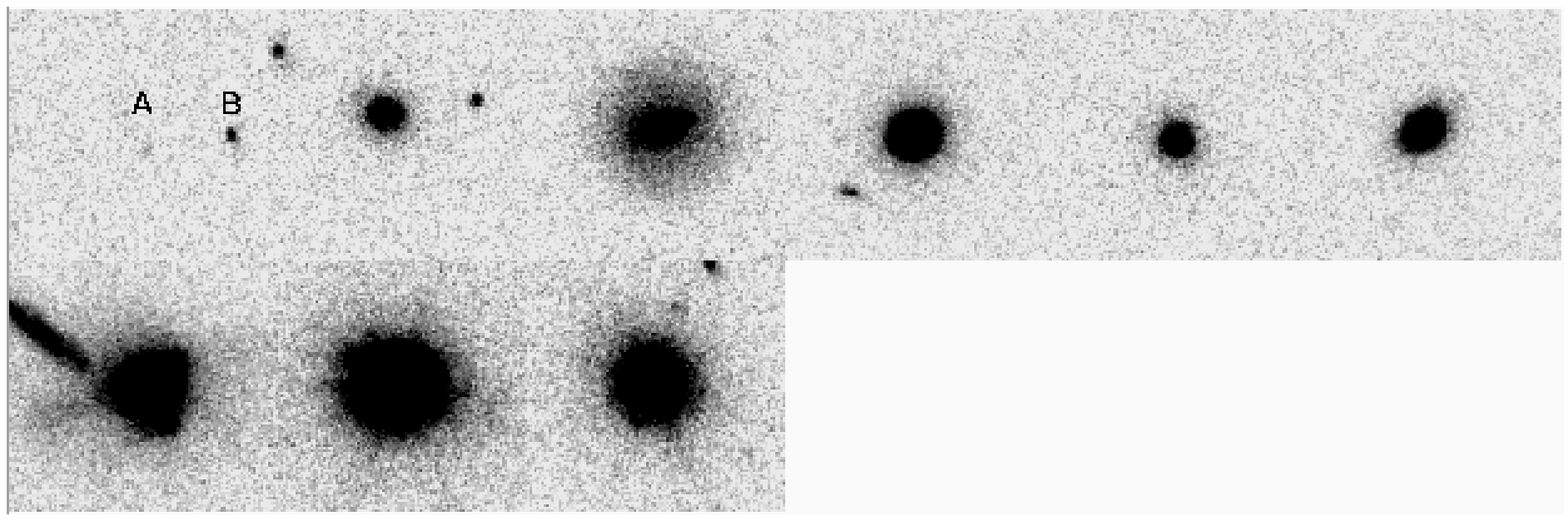}
\vspace*{-0.05in}

\includegraphics[angle=270, width=1in]{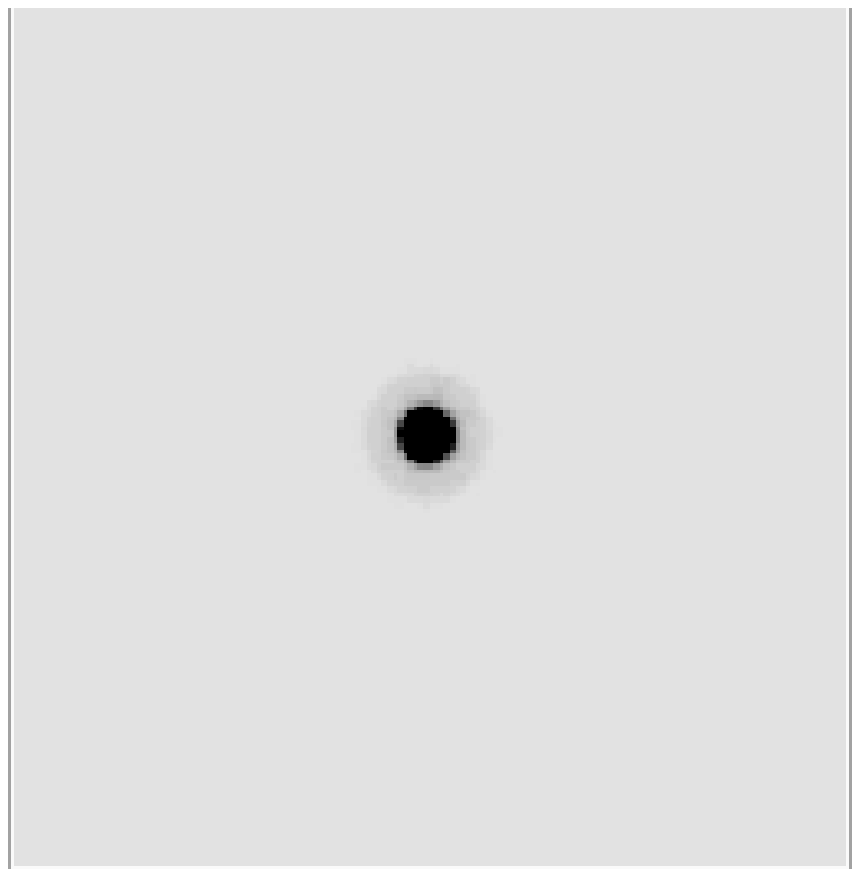}
\figcaption{Postage stamps of the optically--dull AGN (top panel) and 
optically--active AGN (middle panel) at $0.5<z<0.8$. 
Data are ACS/HST i-band images from GOODS \citep{goods_acs};
 each postage stamp is 4.8\arcsec\ on a side.   
Sources are plotted top to bottom, left to right 
in order of increasing absorption--corrected X-ray 
luminosity, as follows: Optically--dull AID: 
139,  196, 157, 83, 212, 298, 
 60,  146,  80, 269, 91, 155,  
276, 126, 176, 134, 22,  48  
161, 247, 227, 44.
Optically--active AID:
241, 229,  103,  117,  195, 179, 88,  66,  177.  
Also shown for comparison is the ACS/HST PSF in i-band, at the same pixel scale
as the images (bottom panel). 
For clarity, we label the two possible counterparts of X-ray source
AID 241 as ``A'' and ``B'', following \citep{szokoly} who identified the
two potential optical counterparts, and obtained a spectrum of B and found
high--excitation emission lines, and thus is presumed to be the X-ray 
counterpart.
}
\label{fig:postage_stamps}
\end{figure*}

\begin{figure}
\figurenum{10}
%
\includegraphics[angle=270,width=3.3in]{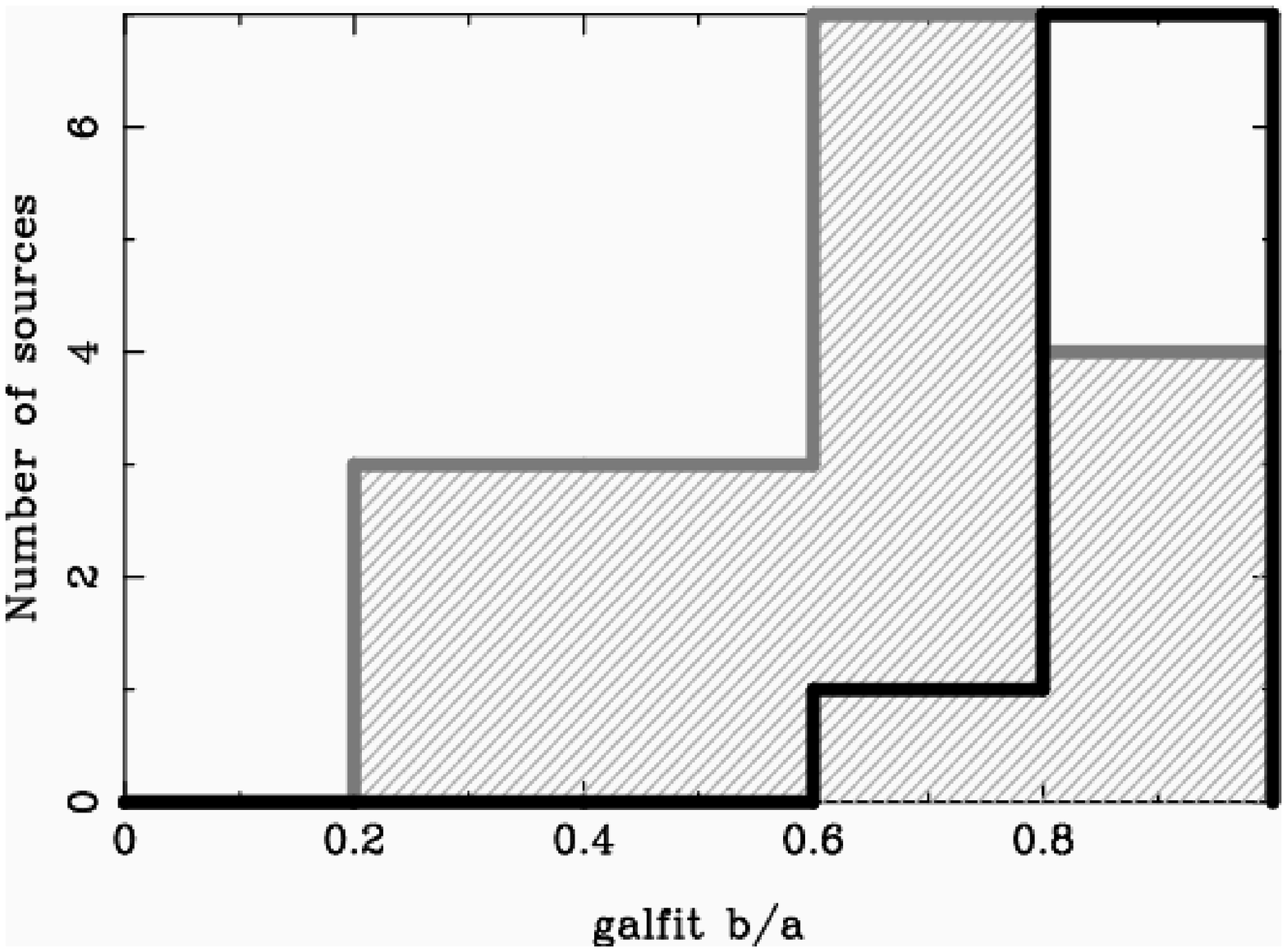}

\includegraphics[angle=270,width=3.3in]{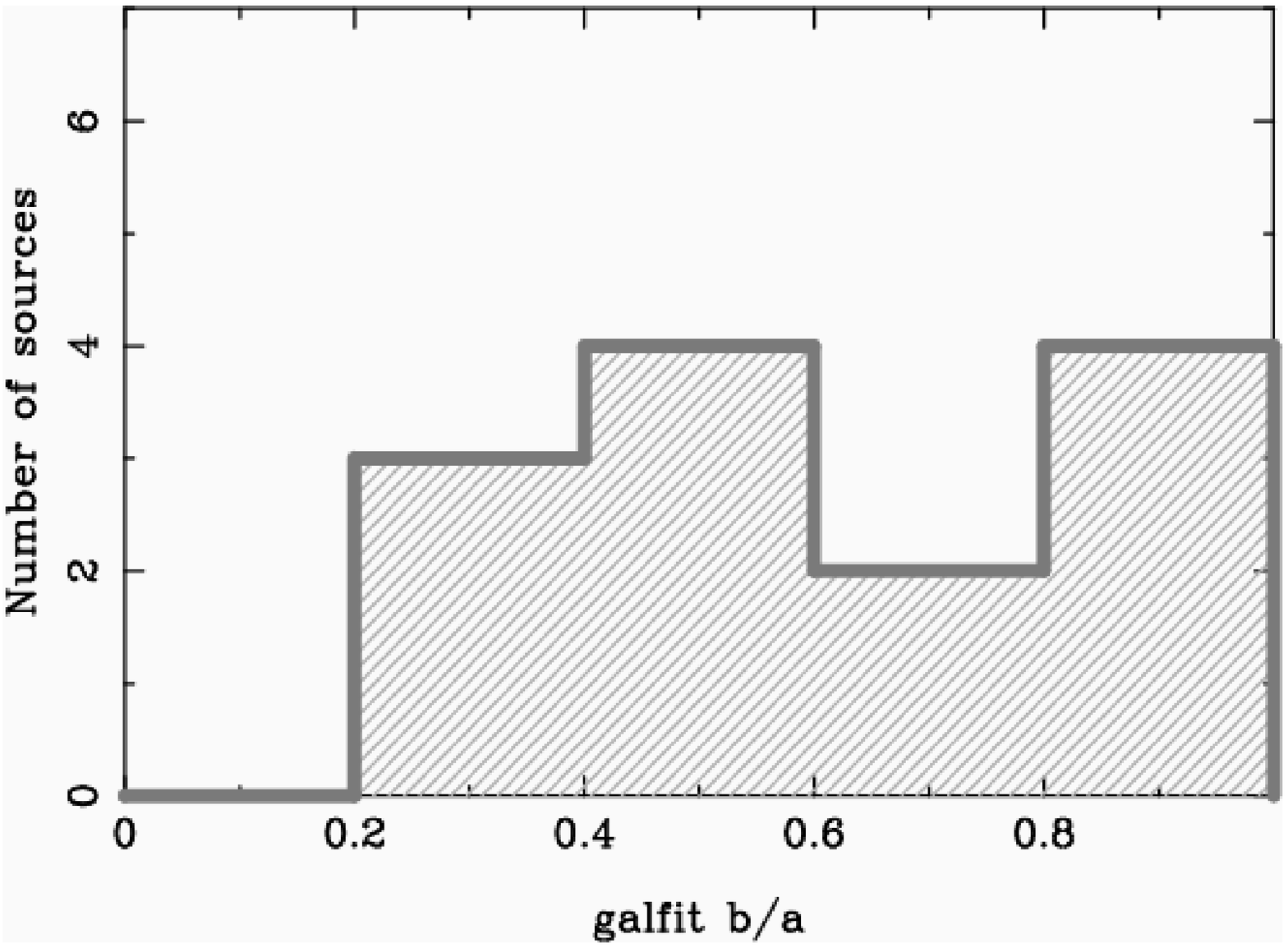}
\figcaption{Axis ratio distributions of the optically--dull 
and optically--active AGN with redshifts $0.5<z<0.8$.  
Axis ratios are from Galfit models.
Upper panel:  The black histogram shows the distribution of the 
optically--active AGN; the hatched histogram shows the distribution 
of the optically--dull AGN.
Lower panel:  Axis ratios for the optically--dull AGN, if a nuclear
point source is added to each. 
}
\label{fig:ba_distrib}
\end{figure}

Tables~\ref{ba_optdull} and \ref{ba_optact} list the 
measured axis ratios for the optically-dull and active samples.  
Figure~\ref{fig:postage_stamps} shows cutouts of the GOODS 
i-band images for the $0.5<z<0.8$ AGN.
Figure~\ref{fig:ba_distrib} plots the axis ratio distributions.
Clearly, the b/a distributions of the two samples are very different: the
optically--active galaxies are round, and the optically-dull galaxies
have a range of b/a.  

Let us explore the differences between these distributions in more detail.  
\textit{All the $0.5<z<0.8$ optically--active sources are less concentrated
than the PSF} (using concentration parameter 
$C= 5 \log r_{80}/r_{20}$  from \citealt{kent85}), and half
are much less concentrated.
Of the optically--active AGN, all but one have b/a$>0.79$ 
\footnote{The one exception is AID 88, whose
complicated morphology may indicate an interaction.}.  Only two
sources, AID 88 and 94, show complicated (and perhaps disturbed)
morphologies.  
Only one source (AID 103) shows any evidence for a possible inclined disk.

The optically--dull AGN are also all resolved.  By contrast with the optically--active
AGN, the optically--dull AGN  
show a range of axis ratio: $0.26 < b/a < 0.89$. 
One source is possibly interacting (AID 196), and there are two irregulars (48 and 60).  
Of the spirals, 3 are highly inclined (AID 44, 134, 146); 
2 are close to face-on (AID 139, 176) 
and 2 have intermediate inclinations (AID 91, 157, 269).
%
The two--sided Kolmogorov-Smirnov test disproves, at $99.7\%$ confidence,
the null hypothesis that the optically-dull and optically--active 
$0.5<z<0.8$ AGN are drawn from the same parent distribution.

We separate out the optically--dull $0.5<z<0.8$ AGN that have low
rest--frame optical-to-X-ray luminosity ratios (no more than 
$\times 6.3$ the Elvis ratio in figure~\ref{fig:mr_lxabscor}c);
they, too, have an axis ratio distribution that differs,
at $99.8\%$ confidence, from that of the optically--active AGN.
These are sources whose $L_R/L_x$ ratios are not high enough for
spectral dilution to be an important cause of optical dullness, and so
it is important to note that their axis ratios are significantly
different from that of the optically--active AGN.
(The optically--dull AGN with high and low $L_R/L_x$ do not 
have significantly different axis ratio distributions, but the
size of each sample is very small.)

Is the optically--active sample biased by nuclear point sources 
toward high axis ratios?
This question arises since 6/9 have broad emission lines, 
and four of these galaxies 
are centrally concentrated, though not as much as a PSF.  
Therefore, for each optically--active AGN in our $0.5<z<0.8$
sample, we measure the \emph{maximum} flux that could come 
from a central point source, assuming no galaxy emission.
The maximal PSF flux is, in the median, $25\%$ of the total
source flux, and corresponds to i$_{AB} = 22.36$.
We then add a PSF of this median brightness, with Poisson noise,
to the centroid of each 
optically--dull source, and repeat the galfit axis ratio measurements.  
Figure~\ref{fig:ba_distrib} shows the result:  
adding artificial point sources to the optically--dull AGN
means that some sources are no longer fit well by any model;
but most sources are still fit acceptably, and the resulting b/a
distribution  does \emph{not} resemble the optically--active 
distribution.  Galfit is able to fit the artificial point source
and still correctly measure the axis ratio of the galaxy.


Thus, we conclude that the host galaxies of optically--dull 
AGN more fully sample the expected range of random inclination angles 
than do optically--active AGN.

\subsection{Additional Axis Ratios from  $z<0.5$ and $z>0.8$ AGN}
\label{sect:morph_otherzs}

Though the $0.5<z<0.8$ redshift sample should be the most robust
against 
morphological bias, we do have
morphological information on the rest of the sample, and we now
briefly examine it.  Since this sample may be biased, it should not be
used to argue that the optically--dull AGN have a more diverse
distribution of axis ratio than the optically--active AGN.  But if the
trends from the previous section are not present in the full redshift 
range, that would weaken the evidence that the trends are real.

Twenty-five optically--active AGN in the CDFS have GOODS ACS imaging and
$z<0.5$ or $z>0.8$.  Most (17) of these are well-fit by a point source
(of which 8 show some faint extended emission after PSF subtraction.)
Thus, only eight sources have morphology information:
one PSF + ring galaxy (with b/a$=0.6$);
two compact but resolved galaxies;
two sources that appear to be undergoing interactions (AID 193 and 214);
and three irregular galaxies.    
Their b/a ratios are listed in table~\ref{ba_optact}.

There are 15 optically-dull AGN with GOODS imaging and $z<0.5$ or
$z>0.8$.  Of these, only 3 are point sources, 4 are irregular or
interacting galaxies, and the remaining 8 sources have b/a ratios from
0.2 to 0.9.  The K-S test finds that the any--redshift optically--dull 
and any--redshift optically--active samples are drawn from different 
populations, at $99.5\%$ significance.
Thus, the higher and lower redshift sources do not contradict the results of the
$0.5<z<0.8$ sample.

\subsection{Column density evidence for extra obscuration}
\label{sect:columns}
\begin{figure}
\figurenum{11}
\includegraphics[angle=270, width=3.3in]{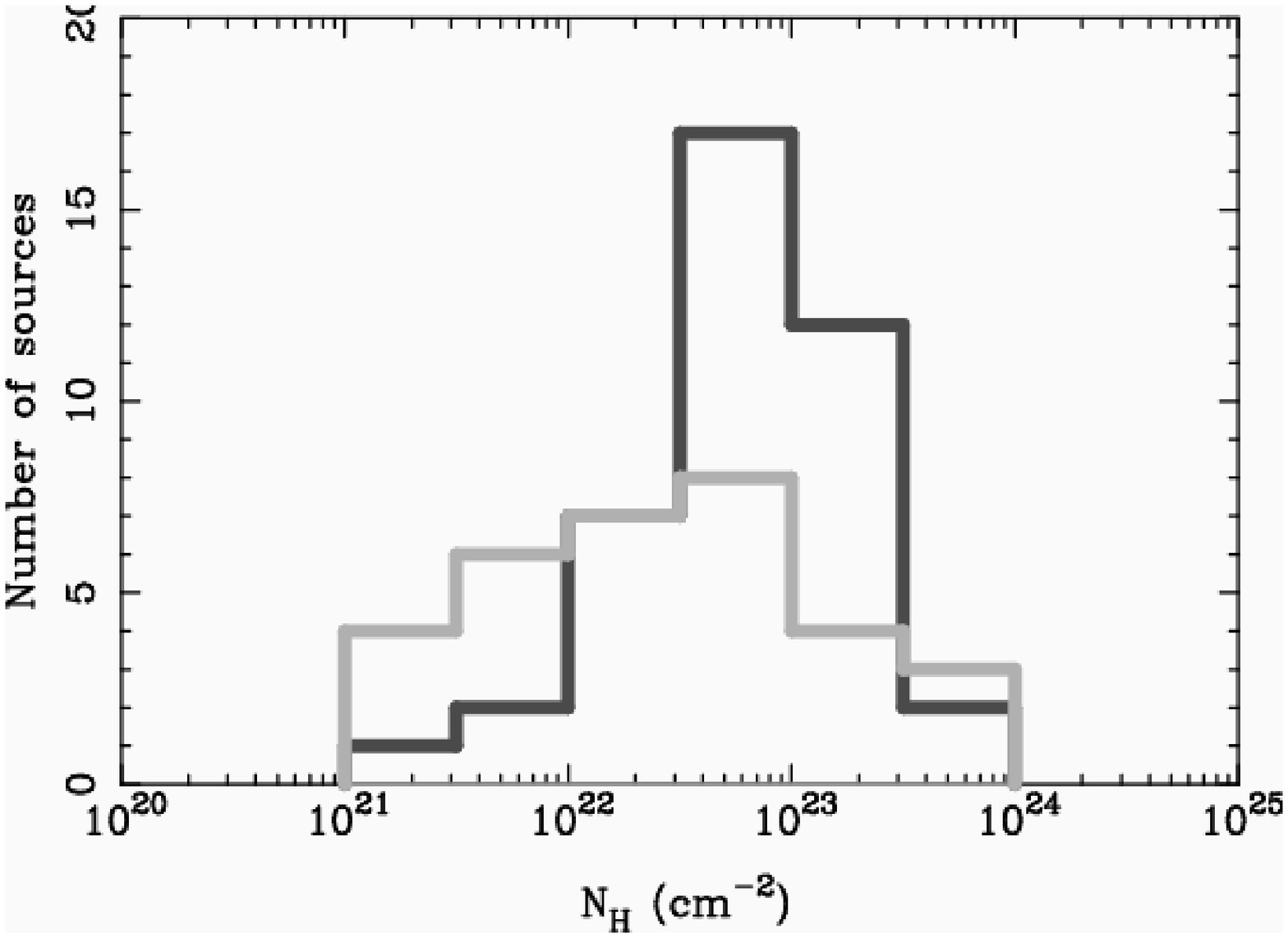}
\figcaption{X-ray column density distributions for the optically--dull 
\textit{(dark histogram)} and optically--active  \textit{(light histogram)} AGN with
absorption--corrected X-ray luminosities below $10^{44}$~\ergss.  A K-S test
shows the two distributions are different at the $95\%$ confidence level; adding back in the
QSOs increases the significance.}
\label{fig:nh-optdull-optact}
\end{figure}

If host galaxy obscuration is a cause of optical dullness, then we might expect
higher X-ray column densities for the optically--dull AGN than the optically--active.
In figure~\ref{fig:nh-optdull-optact} we plot the column density distributions
for the optically--dull and optically active sources with 
absorption--corrected L$_x$ below $10^{44}$~\ergss.  The optically--dull sources
appear to have higher column densities, as expected if the host galaxy contributes
extra extinction.  
This result is marginally significant: the two--sided
Kolmogorov-Smirnov test estimates a $95\%$ probability that the two distributions 
are drawn from different parents.  Adding back in the high--luminosity sources would
increase the significance.
Any such offset is not expected to be large,
given the ROSAT extinction relation 
$A_V = N_H /(1.8\times10^{21})$~mag~cm$^{-2}$ \citep{allens}.

\section{Discussion}
\label{sect:discussion}

We showed in \S\ref{sect:weak-continuua} that optically--dull AGN do not
have unusually weak ionizing continua; we instead found that  
optically--dull AGN have normal Seyfert (UV--powered) mid--infrared emission 
(and thus the optical may be the only wavelength range where these AGN appear odd).
Further, in \S\ref{sect:dilution} we demonstrated  
(by comparing the rest--frame hard X-ray and optical luminosities)
that dilution by stellar continuum is unlikely to explain the optical 
dullness of at least half our sample.   
We thus needed a primary cause for optical dullness in the majority of the
sample.

In \S\ref{sect:morph} we show that obscuration by host galaxies is a 
likely cause of optical dullness for these AGN, which can explain the 
missing emission lines and big blue bumps, and the normal 
Seyfert--like X-ray and mid-IR emission.
We did this by showing that the host galaxies of X-ray--selected AGN
have a range of axis ratio consistent with a wide range of
inclination angle, but that the subset with optical emission lines
have a much narrower range of b/a, which is consistent with occuring
either in nearly face-on galaxies, or in spheroid--dominated galaxies.
Since at $z=1$, the observed
2--8~keV band samples rest--frame 4--16~keV energies, the
Chandra--selected AGN in the deep fields are selected by fairly hard X-rays,
and thus it is not surprising that they, like local hard
X-ray--selected samples, show a large range of axis ratio.  Redshift
will also act to increase the net obscuration in the observed optical
wavelengths of these sources, which should increase the disparity
between optically--selected and hard X-ray--selected samples. 
In addition, bright diagnostic lines like $H\alpha$ are redshifted
out of the optical bands, further increasing the disparity.
Therefore, we propose that the selection effects demonstrated in \citet{mcr95}
for local AGN explain many of the ``optically--dull'' AGN: 
they are missing narrow emission lines due to absorption by extranuclear 
dust in their host galaxies.

This behavior has been modeled by \citet{maiolrieke95} (see their figure~3).
They show that the expected distribution for an unbiased sample of b/a
ratios for randomly oriented disk galaxies is very similar to the
distribution of b/a ratios we measure in figure~\ref{fig:ba_distrib}
using Galfit. They also demonstrate that, when observations are deep
enough to provide a virtually complete sample independent of
orientation, then the distribution of b/a for local Seyfert galaxies
follows this distribution. The local Type 1 galaxies all have b/a near
1, while the galaxies with significantly smaller b/a are observed to
be preferentially types 1.8, 1.9, and 2, thus directly demonstrating
the influence of obscuration in the galaxy disk.

To be precise, the obscuration hypothesis predicts that b/a should be
biased against face-on galaxies.   Such a bias is not seen in 
figure~\ref{fig:ba_distrib}, although the sample size is too small to test
the detailed shape of the distribution.  The
handfull of high b/a sources in the $0.5<z<0.8$ optically--dull sample
are consistent with having spectral dilution explain their optical dullness;
they are centrally concentrated, and have higher--than--average $M_R/L_x$
(at low significance given the small sample size).

\citet{maiolrieke95} also demonstrated that the proportion of missing
AGN---or optically dull X-ray sources---depends critically on the
quality of the spectra used for classification. For the nearby
Seyferts in the Revised Shapley Ames catalog, the bias is almost
absent, while there is a strong bias in the more distant CfA
sample. The small proportion of optically dull X-ray galaxies in
recent studies, as described in \S\ref{sect:lowzhighz}, 
is probably a result of the high quality spectra that can be obtained 
on relatively nearby galaxies.

For standard dust-to-gas ratios, this picture suggests that a
considerable fraction of the column that obscures the soft X-rays may
come from outside the obscuring torus.  This complicates attempts to
interpret X-ray column as a proxy for accretion disk inclination, 
as well as attempts
to use obscured-to-unobscured AGN ratios to estimate torus geometries.
In this picture, X-ray column is not solely a description of the
nuclear obscuration, but of the galactic obscuration as well.
This complex extinction geometry should be included in studies relating
high--redshift AGN to the X-ray background, for example.  

Also, this picture may partially explain the lack of dependence of the
hard X-ray to $24$~\micron\ flux ratio on the X-ray hardness ratio (a
proxy for column density).  If many X-ray sources are partially
obscured by gas and dust far away from the AGN, then the dust will not
be heated sufficiently to emit in the mid--infrared.  Thus, having the
obscuration take place in the host galaxy as well as the torus will
dilute the signatures of torus obscuration that were searched for, but
not seen, in \citet{rigbyx24} and \citet{lutz}.

\section{Conclusions}
\label{sect:conclusions}

We have investigated the column densities, X-ray and IR luminosities,
and morphologies of 45 X-ray--selected AGN in the CDFS
that lack optical AGN emission lines (and are thus termed 
``optically--dull AGN''.)

We test whether these sources are low--luminosity AGN in very luminous
galaxies; this would support the hypothesis that the AGN emission
lines are drowned out by bright galactic continua (``dilution'').
Fifty-six percent of our sample have rest frame R-band luminosities no more
than $\sim6$ times larger than that expected for the AGN (scaling from the
X-ray luminosity), and thus \emph{we conclude that dilution is not the
primary cause of optical dullness} for at least half the sample.
This should be contrasted with
the bright $z\sim0.2$ optically--dull AGN of \citet{comastri}, which have
high optical--to--X-ray flux ratios that make dilution likely.

About half of the local weak--line AGN have column densities much
lower than those of the CDFS sample ($\log N_H < 22$~\cmsq).  In
addition, it has been shown that dilution can account for the
optical dullness of many of these low-redshift galaxies.
Thus, they are not true analogues to the $z\sim1$ population.

Optically--dull AGN
have the normal mid--infrared emission we expect from Seyfert
galaxies; in short, they look like AGN at $24$~\micron.  
Since AGN IR emission is powered by UV continua, it is likely that
they have normal amounts of UV emission.

We test whether the morphologies of optically--dull AGN are atypical.
The optically--dull AGN have host galaxies with a large range of inclination angle,
whereas the optically--active AGN hosts are nearly--face-on spirals or
spheroids (and thus should have less dust extinction).
  From this, we conclude that X-ray--selected AGN in deep
fields are selected fairly independently of their inclination angle,
but that only the most face-on or spherical show optical emission
lines.  In the rest, extranuclear dust in the host galaxy may obscure
the narrow-line regions.  This scenario is consistent with samples of
Seyferts in the local universe, where hard X-ray and mid-IR--selected
samples have unbiased b/a distributions, but optically--selected
samples of Seyfert 1 and 2 AGN are systematically biased against
inclined disk galaxies.

Thus, part of the column density that obscures the soft X-rays may come from
the host galaxy, outside the obscuring torus.  This complicates
using the X-ray column to infer torus properties.  It may also partially 
explain why the mid--infrared to X-ray luminosity ratio does not depend
on column density in AGN.

Thus, we conclude that host galaxy obscuration is the primary cause of optical
dullness, with spectral dilution a likely contributor for sources with high
optical to absorption--corrected X-ray luminosity ratios.

\vspace*{0.1in}
\acknowledgements
We thank G. Szokoly for kindly providing electronic versions of the 
VLT spectra in advance of online publication, and Steve Willner for 
comments that improved the manuscript.
This work is based in part on observations made with \emph{Spitzer}, 
which is operated by the Jet Propulsion Laboratory, 
California Institute of Technology under NASA contract 1407. 
Support for this work was provided by NASA through Contract Number 960785 
issued by JPL/Caltech.

\clearpage
\begin{landscape}
\pagestyle{empty}
\setlength{\hoffset}{-10mm}
\begin{deluxetable}{llllllllllllllllll}
\tabletypesize{\tiny}
\tablecolumns{18}
\tablewidth{0pc}
\tablecaption{Optically--Dull AGN.\label{tab:sample}}
\tablehead{
\colhead{AID} &  \colhead{SID} &  \colhead{A-RA} & 
\colhead{A-DEC} &  \colhead{S-RA} &  \colhead{S-DEC} & \colhead{off} & \colhead{2-8 keV flux} &
\colhead{log f(H)$/$f(S)} &  \colhead{f$_{\nu}$ 24~\micron} & 
\colhead{L$_x$} & \colhead{corr L$_x$} & \colhead{est N(H)} &
\colhead{X class} & \colhead{Opt} &  \colhead{Q} & \colhead{z} & \colhead{SED} 
}
\startdata
$ 12$ & $121$  & $52.96316$ & $-27.84766$ &   $52.96316$ & $-27.84766$ & $0.0$ & $1.4 (\pm0.4)$ E-15 & $0.47^{+0.13}_{-0.19}$ & $<96$        &  1.8E42 & 2.6E42 &   1.9E22 &   AGN-2 & LEX &  3  & 0.674  &  O\\
$ 16$ & $ 76$  & $52.96875$ & $-27.83819$ &   $52.96870$ & $-27.83822$ & $0.2$ & $8.6 (\pm0.7)$ E-15 & $0.73^{+0.05}_{-0.05}$ & $87\pm30$    &  9.3E43 & 3.9E44 &   3.2E23 &   QSO-2 & LEX &  1  & 2.394  &  F\\
$ 28$ & $ 73$  & $52.99208$ & $-27.80944$ &   $52.99175$ & $-27.80963$ & $1.3$ & $4.8 (\pm0.5)$ E-15 & $0.43^{+0.05}_{-0.05}$ & $82\pm9$     &  7.9E42 & 1.1E43 &   2.0E22 &   AGN-1 & LEX &  3  & 0.734  &  O\\
$ 44$ & $ 66$  & $53.01520$ & $-27.76769$ &   $53.01529$ & $-27.76772$ & $0.3$ & $1.7 (\pm0.1)$ E-14 & $1.44^{+0.06}_{-0.07}$ & $126\pm30$   &  7.5E42 & 2.2E43 &   1.1E23 &   AGN-2 & LEX &  3  & 0.574  &  O\\
$ 48$ & $267$  & $53.02037$ & $-27.69100$ &   $53.02029$ & $-27.69100$ & $0.3$ & $5.7 (\pm1)$ E-15   & $> 1.38$ 	      & $91\pm10$    &  3.8E42 & 1.3E43 & $>1.4$E23&   AGN-2 & LEX &  1  & 0.720  &  F\\
$ 60$ & $155$  & $53.03295$ & $-27.71091$ &   $53.03325$ & $-27.71094$ & $0.9$ & $1.2 (\pm0.4)$ E-15 & $0.71^{+0.15}_{-0.22}$ & $520\pm20$   &  8.3E41 & 1.4E42 &   3.3E22 &   AGN-2 & LEX &  3  & 0.545  &  R\\
$ 65$ & $538$  & $53.03541$ & $-27.78016$ &   $53.03562$ & $-27.78011$ & $0.7$ & $1.1 (\pm0.3)$ E-15 & $> 1.10$ 	      & $990\pm80$   &  2.1E41 & 3.4E41 & $>4.6$E22&   AGN-2 & LEX &  3  & 0.310  &  O\\
$ 80$ & $535$  & $53.04758$ & $-27.78050$ &   $53.04758$ & $-27.78055$ & $0.2$ & $1.3 (\pm0.3)$ E-15 & $0.76^{+0.12}_{-0.16}$ & $<80$        &  9.8E41 & 1.7E42 &   3.7E22 &   AGN-2 & LEX &  3  & 0.575  &  O\\
$ 83$ & $534$  & $53.05062$ & $-27.75825$ &   $53.05087$ & $-27.75838$ & $0.9$ & $<4.8$ E-16         & $< 0.87$ 	      & $<80$        & $<6.9$E41 & $<9.3$E41 &  $<6.2$E22&  AGN-2 & LEX &  3  & 0.676  &  O\\
$ 84$ & $149$  & $53.05104$ & $-27.77247$ &   $53.05125$ & $-27.77272$ & $1.1$ & $1.3 (\pm0.3)$ E-15 & $0.94^{+0.14}_{-0.20}$ & $239\pm40$   &  2.5E42 & 7.3E42 &   1.2E23 &   AGN-2 & LEX &  1  & 1.033  &  Y\\ 
$ 87$ & $156$  & $53.05504$ & $-27.92469$ &   $53.05512$ & $-27.92463$ & $0.3$ & $7.0 (\pm0.9)$ E-15 & $> 1.77$ 	      & $116\pm10$   &  5.6E42 & 5.5E43 & $>4.2$E23&   AGN-2 & ABS &  3  & 1.185  &  I\\
$ 90$ & $600$  & $53.05762$ & $-27.75711$ &   $53.05783$ & $-27.75736$ & $1.1$ & $8.1 (\pm3)$ E-16   & $> 1.21$ 	      & $<80$        &  1.6E42 & 8.4E42 & $>2.7$E23&   AGN-2 & LEX &  3  & 1.327  &  I\\
$ 91$ & $266$  & $53.05779$ & $-27.71338$ &   $53.05775$ & $-27.71361$ & $0.8$ & $1.2 (\pm0.4)$ E-15 & $> 1.22$ 	      & $289\pm40$   &  9.9E41 & 3.0E42 & $>1.2$E23&   AGN-2 & LEX &  3  & 0.735  &  Y\\
$126$ & $ 50$  & $53.07912$ & $-27.79872$ &   $53.07916$ & $-27.79872$ & $0.1$ & $2.3 (\pm0.4)$ E-15 & $0.86^{+0.09}_{-0.11}$ & $<80$        &  2.2E42 & 4.4E42 &   5.4E22 &   AGN-2 & ABS &  1  & 0.670  &  F\\
$129$ & $525$  & $53.08254$ & $-27.68975$ &   $53.08250$ & $-27.68966$ & $0.3$ & $<9.4$ E-16         & $< 0.62$  	      & $3018\pm90$  & $<1.1$E41 & $<1.4$E41 & $<1.6$E22&   AGN-2 & LEX &  3  & 0.229  &  O\\
$131$ & $253$  & $53.08362$ & $-27.74638$ &   $53.08366$ & $-27.74644$ & $0.2$ & $4.2 (\pm0.5)$ E-15 & $1.55^{+0.12}_{-0.16}$ & $333\pm30$   &  1.3E42 & 3.6E42 &   1.2E23 &   AGN-2 & LEX &  1  & 0.481  &  F?\\
$134$ & $151$  & $53.08529$ & $-27.79233$ &   $53.08533$ & $-27.79230$ & $0.2$ & $7.6 (\pm0.7)$ E-15 & $2.06^{+0.15}_{-0.22}$ & $65\pm20$    &  2.2E42 & 1.1E43 &   2.2E23 &   AGN-2 & LEX &  3  & 0.604  &  O\\
$139$ & $602$  & $53.09141$ & $-27.78219$ &   $53.09158$ & $-27.78216$ & $0.5$ & $3.3 (\pm1)$ E-16   & $> 0.82$ 	      & $222\pm70$   &  4.7E41 & 6.3E41 & $>5.0$E22&   AGN-2 & ABS &  3  & 0.668  &  O\\
$146$ & $188$  & $53.09400$ & $-27.83050$ &   $53.09400$ & $-27.83050$ & $0.0$ & $6.3 (\pm2)$ E-16   & $0.86^{+0.16}_{-0.25}$ & $69\pm10$    &  7.0E41 & 1.5E42 &   6.4E22 &   AGN-2 & LEX &  3  & 0.734  &  F\\
$155$ & $ 49$  & $53.10095$ & $-27.69066$ &   $53.10104$ & $-27.69069$ & $0.3$ & $3.1 (\pm0.4)$ E-15 & $0.23^{+0.06}_{-0.07}$ & $68\pm9$     &  2.9E42 & 3.4E42 &   5.8E21 &   AGN-1 & LEX &  3  & 0.534  &  O\\
$157$ & $598$  & $53.10283$ & $-27.90325$ &   $53.10283$ & $-27.90322$ & $0.1$ & $<5.5$ E-16         & \nodata 		      & $<80$        & $<6.4$E41 & $<8.5$E41 &  \nodata &   AGN-2 & ABS &  3  & 0.617  &  I\\
$161$ & $ 47$  & $53.10416$ & $-27.68383$ &   $53.10404$ & $-27.68377$ & $0.5$ & $5.5 (\pm0.8)$ E-15 & $1.25^{+0.11}_{-0.15}$ & $155\pm10$   &  4.2E42 & 1.3E43 &   1.3E23 &   AGN-2 & LEX &  3 &  0.733 &   Y\\
$162$ & $260$  & $53.10466$ & $-27.84538$ &   $53.10462$ & $-27.84536$ & $0.2$ & $1.3 (\pm0.3)$ E-15 & $1.24^{+0.16}_{-0.24}$ & $30\pm10$    &  1.8E42 & 7.7E42 &   1.9E23 &   AGN-2 & LEX &  3 &  1.043 &   F\\
$164$ & $150$  & $53.10487$ & $-27.91377$ &   $53.10483$ & $-27.91391$ & $0.5$ & $3.9 (\pm0.6)$ E-15 & $> 1.67$ 	      & $108\pm30$   &  3.3E42 & 2.5E43 & $>3.4$E23&   AGN-2 & ABS &  3 &  1.090 &   O\\
$166$ & $ 45$  & $53.10700$ & $-27.71827$ &   $53.10700$ & $-27.71825$ & $0.1$ & $4.9 (\pm0.5)$ E-15 & $0.66^{+0.05}_{-0.06}$ & $480\pm50$   &  5.7E43 & 2.0E44 &   2.4E23 &   QSO-2 & LEX &  1 &  2.291 &   R\\
$171$ & $519$  & $53.10770$ & $-27.91875$ &   $53.10779$ & $-27.91844$ & $1.1$ & $<7.8$ E-16         & $< 0.64$ 	      & $63\pm10$    & $<2.1$E42 & $<4.3$E42 & $<5.8$E22&   AGN-2 & LEX &  3 &  1.034 &   F\\
$176$ & $ 43$  & $53.11150$ & $-27.69600$ &   $53.11150$ & $-27.69600$ & $0.0$ & $3.7 (\pm0.5)$ E-15 & $0.78^{+0.07}_{-0.08}$ & $385\pm40$   &  4.4E42 & 8.7E42 &   5.4E22 &   AGN-2 & LEX &  3 &  0.734 &   I\\
$196$ & $516$  & $53.13058$ & $-27.79027$ &   $53.13083$ & $-27.79019$ & $0.9$ & $<3.7$ E-16         & $< 0.71$ 	      & $116\pm8$    & $<5.1$E41 & $<6.8$E41 & $<3.9$E22&   AGN-1 & LEX &  3 &  0.665 &   F\\
$212$ & $512$  & $53.14329$ & $-27.73061$ &   $53.14312$ & $-27.73063$ & $0.5$ & $6.8 (\pm2)$ E-16   & $0.59^{+0.15}_{-0.23}$ & $<80$        &  7.9E41 & 1.3E42 &   2.8E22 &   AGN-2 & LEX &  3 &  0.668 &   O\\
$216$ & $171$  & $53.14629$ & $-27.73627$ &   $53.14662$ & $-27.73655$ & $1.5$ & $<4.3$ E-16         & $< 0.54$ 	      & $202\pm40$   & $<1.0$E42 & $<1.4$E42 & $<3.3$E22&   AGN-2 & LEX &  1 &  0.839 &   F\\ 
$220$ & $190$  & $53.14933$ & $-27.68333$ &   $53.14941$ & $-27.68322$ & $0.5$ & $5.4 (\pm0.8)$ E-15 & $> 1.62$ 	      & $<80$        &  3.0E42 & 1.3E43 & $>1.9$E23&   AGN-2 & LEX &  3 &  0.735 &   O\\
$221$ & $100$  & $53.14991$ & $-27.81397$ &   $53.14991$ & $-27.81400$ & $0.1$ & $3.1 (\pm1.3)$ E-16 & $0.17^{+0.16}_{-0.27}$ & $43\pm9$     &  2.6E42 & 3.1E42 &   9.9E21 &   AGN-1 & LEX &  1 &  1.309 &   F\\
$227$ & $ 33$  & $53.15295$ & $-27.73508$ &   $53.15329$ & $-27.73533$ & $1.4$ & $8.6 (\pm0.6)$ E-15 & $0.33^{+0.03}_{-0.03}$ & $94\pm20$    &  1.2E43 & 1.6E43 &   1.1E22 &   AGN-1 & LEX &  3 &  0.665 &   I\\
$247$ & $ 25$  & $53.17016$ & $-27.92961$ &   $53.17020$ & $-27.92972$ & $0.4$ & $9.3 (\pm0.8)$ E-15 & $1.24^{+0.06}_{-0.07}$ & $661\pm50$   &  5.6E42 & 1.5E43 &   9.5E22 &   AGN-2 & ABS &  0.5 &0.625 &   R\\
$259$ & $132$  & $53.18358$ & $-27.91502$ &   $53.18337$ & $-27.91502$ & $0.7$ & $8.4 (\pm3)$ E-16   & $0.55^{+0.15}_{-0.24}$ & $<80$        &  1.9E42 & 3.4E42 &   3.5E22 &   AGN-2 & LEX &  1 &  0.908 &   F?\\
$264$ & $ 85$  & $53.18587$ & $-27.80991$ &   $53.18583$ & $-27.80997$ & $0.2$ & $1.3 (\pm0.3)$ E-15 & $0.45^{+0.09}_{-0.11}$ & $<80$        &  2.9E43 & 7.1E43 &   1.5E23 &   QSO-1 & LEX &  1 &  2.593 &   F\\
$269$ & $170$  & $53.19329$ & $-27.90383$ &   $53.19337$ & $-27.90388$ & $0.3$ & $1.4 (\pm0.4)$ E-15 & $0.79^{+0.13}_{-0.18}$ & $218\pm30$   &  1.3E42 & 2.6E42 &   4.7E22 &   AGN-2 & ABS &  3 &  0.664 &   O\\
$271$ & $252$  & $53.19575$ & $-27.72950$ &   $53.19595$ & $-27.72966$ & $0.9$ & $2.9 (\pm0.5)$ E-15 & $1.39^{+0.15}_{-0.22}$ & $<80$        &  3.8E42 & 2.2E43 &   2.9E23 &   AGN-2 & LEX &  3 &  1.178 &   Y\\
$274$ & $ 18$  & $53.19941$ & $-27.70911$ &   $53.19958$ & $-27.70911$ & $0.5$ & $3.0 (\pm0.1)$ E-14 & $0.57^{+0.02}_{-0.02}$ & $1114\pm10$  &  7.7E43 & 1.4E44 &   4.5E22 &   QSO-1 & LEX &  3 &  0.979 &   F?\\
$276$ & $184$  & $53.20075$ & $-27.88236$ &   $53.20075$ & $-27.88244$ & $0.3$ & $2.0 (\pm0.4)$ E-15 & $1.11^{+0.14}_{-0.21}$ & $156\pm30$   &  1.5E42 & 3.7E42 &   8.6E22 &   AGN-2 & ABS &  3 &  0.667 &   O\\
$285$ & $242$  & $53.21595$ & $-27.70802$ &   $53.21600$ & $-27.70822$ & $0.7$ & $<1.5$ E-15         & $< 0.62$ 	      & $437\pm30$   & $<5.3$E42 & $<8.1$E42 & $<5.5$E22 &   AGN-1 & LEX &  3 &  1.027 &   Y\\
$298$ & $110$  & $53.24429$ & $-27.77569$ &   $53.24420$ & $-27.77555$ & $0.6$ & $<8.3$ E-16         & $< 0.43$     	      & $<80$        & $<9.4$E41 & $<1.3$E42 & $<1.5$E22 &   AGN-1 & LEX &  3 &  0.622 &   F\\
$303$ & $ 12$  & $53.24858$ & $-27.84172$ &   $53.24870$ & $-27.84177$ & $0.5$ & $4.3 (\pm0.5)$ E-15 & $0.21^{+0.05}_{-0.05}$ & $<80$        &  7.4E41 & 7.9E41 &   3.0E21 &   AGN-1 & ABS &  3 &  0.251 &   O\\
$304$ & $ 10$  & $53.24904$ & $-27.77394$ &   $53.24904$ & $-27.77400$ & $0.2$ & $9.2 (\pm0.7)$ E-15 & $0.97^{+0.05}_{-0.06}$ & $108\pm30$   &  3.3E42 & 5.8E42 &   4.4E22 &   AGN-2 & LEX &  3 &  0.424 &   O\\
$324$ & $176$  & $53.28829$ & $-27.74688$ &   $53.28854$ & $-27.74722$ & $1.4$ & $2.7 (\pm0.8)$ E-15 & $0.44^{+0.12}_{-0.17}$ & $125\pm30$   &  5.2E42 & 7.7E42 &   2.1E22 &   AGN-1 & LEX &  3 &  0.786 &   Y\\
\enddata
\tablecomments{
Columns: (1) X-ray source identification from \citet{alex}.   
(2) Source ID from \citet{szokoly}, which is identical to the \citet{giacconi} XID.   
(3)--(4) RA and DEC (J2000) of the X-ray source, from \citet{alex}  
(5)--(6) RA and DEC (J2000) of the optical counterpart, from \citet{szokoly}.
(7) Offset between X-ray and optical coordinates, in arcseconds.   
(8)--2-8 keV X-ray flux in \cgsflux, from \citet{alex}.  
(9) log of the 2--8kev / 0.5--2 keV flux ratio.
(10) observed MIPS f$_{\nu}$ (24~\micron) in $\mu$Jy.
(11) rest-frame 2--8 keV X-ray luminosity, using the photon index inferred from the 
2--8kev / 0.5--2 keV flux ratio, and the 2--8 keV flux for normalization.
(12)  absorption-corrected rest-frame 2--8 keV X-ray luminosity, extrapolated from 
the observed 2--8 keV flux, assuming an intrisic photon index $\Gamma=2$ 
and concordence cosmology ($\Omega_{m} = 0.27$, $\Omega_{\Lambda} = 0.73$, $h=0.72$).
(13) Estimated column density:  the column density which would produce the observed 
2--8kev / 0.5--2 keV flux ratio, given an intrinsic $\Gamma=2$ power-law spectrum.
(14)--(17)  X-ray source classification, optical source classification, redshift quality flag, 
and redshift, all from \citet{szokoly}.
(18)  SED classification, this paper:   O=old stellar population;  Y=young stellar population; I=intermediate stellar population;  
F=flat in $\nu$f$_{\nu}$;  R=rising in $\nu$f$_{\nu}$. 
}

\end{deluxetable}
\clearpage
\end{landscape}
\LongTables
\begin{landscape}
\setlength{\hoffset}{-10mm}
\begin{deluxetable}{lllllllllllllllll}
\tabletypesize{\tiny}
\tablecolumns{17}
\tablewidth{0pc}
\tablecaption{Optically--Active AGN.\label{tab:sample-optact}}
\tablehead{
\colhead{AID} &  \colhead{SID} &  \colhead{A-RA} & 
\colhead{A-DEC} &  \colhead{S-RA} &  \colhead{S-DEC} & \colhead{off} & \colhead{2-8 keV flux} &
\colhead{log f(H)$/$f(S)} &  \colhead{f$_{\nu}$ 24~\micron} & 
\colhead{L$_x$} & \colhead{corr L$_x$} & \colhead{est N(H)} &
\colhead{X class} & \colhead{Opt} &  \colhead{Q} & \colhead{z}
}
\startdata
   $5$  & $238$ & $52.94991$  & $-27.84597$  & $52.94987$  & $-27.84597$  & $0.1$  & 3.4($\pm0.5$)E-15  & $0.14^{-0.07}_{+0.08}$  & $<80$       & 1.8E43  & 2.1E43  & 6.5E21  & AGN-1  & BLAGN  & $3$  & $1.065$\\
  $14$  & $112a$ & $52.96658$ & $-27.89075$  & $52.96641$  & $-27.89094$  & $0.9$  & $<$1.3E-15       & $<0.67$                 & $340\pm10$ & $<2.3$E43 & $<9.9$E43 & $<3.8$E23  & QSO-2  & HEX  & $3$  & $2.940$\\
  $18$  & $230$  & $52.97312$ & $-27.81197$  & $52.97312$  & $-27.81197$  & $0.0$  & $<$8.0E-16       & $<0.33$                 & $<80$      & $<1.6$E43 & $<2.9$E43 & $<6.4$E22 & AGN-1  & BLAGN  & $3$  & $2.185$\\
 $22$  & $75$   & $52.98079$  & $-27.91325$  & $52.98075$  & $-27.91344$  & $0.7$  & 1.7($\pm0.2$)E-14  & $1.13^{-0.08}_{+0.10}$  & $73\pm13$   & 1.5E43  & 4.2E43  & 1.1E23  & AGN-2  & HEX  & $3$  & $0.737$\\
  $34$  & $71$  & $53.00158$  & $-27.72211$  & $53.00145$  & $-27.72211$  & $0.4$  & 7.0($\pm0.6$)E-15  & $0.29^{-0.04}_{+0.05}$  & $100\pm10$  & 2.9E43  & 3.9E43  & 1.6E22  & AGN-1  & BLAGN  & $3$  & $1.037$\\
  $39$  & $68$  & $53.00662$  & $-27.72419$  & $53.00658$  & $-27.72416$  & $0.2$  & 5.6($\pm0.5$)E-15  & $0.31^{-0.04}_{+0.05}$  & $<80$       & 1.9E44  & 3.4E44  & 8.8E22  & QSO-1  & BLAGN  & $3$  & $2.726$\\
  $41$  & $67$  & $53.01025$  & $-27.76675$  & $53.01029$  & $-27.76677$  & $0.2$  & 8.6($\pm0.6$)E-15  & $0.30^{-0.04}_{+0.04}$  & $200\pm50$  & 9.5E43  & 1.4E44  & 3.0E22  & QSO-1  & BLAGN  & $3$  & $1.616$\\
  $43$  & $117$ & $53.01262$  & $-27.74738$  & $53.01270$  & $-27.74727$  & $0.5$  & 9.0($\pm3$)E-16    & $0.12^{-0.12}_{+0.16}$  & $<80$       & 3.9E43  & 4.9E43  & 2.1E22  & QSO-1  & HEX  & $3$  & $2.573$\\
  $63$  & $89$  & $53.03433$  & $-27.69822$  & $53.03450$  & $-27.69822$  & $0.5$  & $<$8.1E-16       & $<0.23$                 & $<80$    & $<2.6$E43 &$<4.0$E43 & $<4.5$E22  & AGN-1  & BLAGN  & $3$  & $2.47$\\
  $66$  & $63$  & $53.03608$  & $-27.79288$  & $53.03616$  & $-27.79288$  & $0.3$  & 6.8($\pm0.1$)E-14  & $0.18^{-0.01}_{+0.01}$  & $3300\pm100$ & 6.8E43  & 7.6E43  & 4.3E21  & QSO-1  & BLAGN  & $3$  & $0.543$\\
  $68$  & $62$  & $53.03937$  & $-27.80188$  & $53.03941$  & $-27.80188$  & $0.1$  & 6.0($\pm0.5$)E-15  & $0.73^{-0.05}_{+0.05}$  & $840\pm60$  & 8.2E43  & 3.9E44  & 4.1E23  & QSO-2  & BLAGN  & $3$  & $2.810$\\
  $76$  & $60$  & $53.04545$  & $-27.73752$  & $53.04550$  & $-27.73755$  & $0.2$  & 9.8($\pm0.6$)E-15  & $0.20^{-0.03}_{+0.03}$  & $290\pm30$  & 1.2E44  & 1.6E44  & 1.8E22  & QSO-1  & BLAGN  & $3$  & $1.615$\\
 $86$  & $57$   & $53.05395$  & $-27.87686$  & $53.05400$  & $-27.87691$  & $0.2$  & 4.7($\pm0.5$)E-15  & $0.75^{-0.06}_{+0.07}$  & $49\pm9$    & 5.4E43  & 2.5E44  & 3.8E23  & QSO-2  & HEX  & $3$  & $2.562$\\
 $88$  & $56a$  & $53.05516$  & $-27.71136$  & $53.05516$  & $-27.71141$  & $0.2$  & 1.9($\pm0.1$)E-14  & $0.88^{-0.03}_{+0.03}$  & $700\pm20$  & 1.4E43  & 2.8E43  & 5.3E22  & AGN-2  & HEX  & $3$  & $0.605$\\
 $94$  & $55$   & $53.05837$  & $-27.85022$  & $53.05841$  & $-27.85025$  & $0.2$  & 9.1($\pm0.7$)E-15  & $1.10^{-0.05}_{+0.06}$  & $140\pm90$  & 2.7E41  & 3.4E41  & 3.0E22  & AGN-2  & HEX  & $3$  & $0.122$\\
 $96$  & $531$  & $53.06012$  & $-27.85305$  & $53.06016$  & $-27.85302$  & $0.2$  & 1.2($\pm0.3$)E-15  & $1.28^{-0.17}_{+0.29}$  & $100\pm10$  & 2.6E42  & 1.9E43  & 3.8E23  & AGN-2  & HEX  & $3$  & $1.544$\\
 $98$  & $54$   & $53.06070$  & $-27.90600$  & $53.06087$  & $-27.90575$  & $1.0$  & 2.9($\pm0.4$)E-15  & $0.66^{-0.08}_{+0.09}$  & $<80$       & 4.0E43  & 1.5E44  & 2.9E23  & QSO-2  & HEX  & $3$  & $2.561$\\
 $103$  & $53$  & $53.06245$  & $-27.85755$  & $53.06245$  & $-27.85755$  & $0.0$  & 3.4($\pm0.4$)E-15  & $0.29^{-0.05}_{+0.06}$  & $75\pm25$    & 5.2E42  & 6.5E42  & 1.0E22  & AGN-1  & BLAGN  & $3$  & $0.675$\\
 $109$  & $206$ & $53.06754$  & $-27.65844$  & $53.06750$  & $-27.65850$  & $0.2$  & 1.8($\pm0.1$)E-14  & $0.15^{-0.03}_{+0.03}$ & $1350\pm75$  & 1.5E44  & 1.9E44  & 8.9E21  & QSO-1  & BLAGN  & $3$  & $1.324$\\
 $117$  & $52$  & $53.07141$  & $-27.71758$  & $53.07145$  & $-27.71761$  & $0.2$  & 6.4($\pm0.5$)E-15  & $0.15^{-0.04}_{+0.04}$  & $1370\pm90$ & 7.3E42  & 8.1E42  & 3.2E21  & AGN-1  & BLAGN  & $3$  & $0.569$\\
 $122$  & $87$  & $53.07600$  & $-27.87819$  & $53.07604$  & $-27.87816$  & $0.2$  & 6.3($\pm2$)E-16    & $0.24^{-0.13}_{+0.19}$  & $<80$       & 2.6E43  & 4.2E43  & 6.4E22  & AGN-1  & BLAGN  & $3$  & $2.801$\\
 $123$  & $153$ & $53.07641$  & $-27.84866$  & $53.07645$  & $-27.84869$  & $0.2$  & 7.4($\pm0.7$)E-15  & $1.65^{-0.09}_{+0.12}$  & $130\pm14$  & 8.9E42  & 1.1E44  & 5.9E23  & AGN-2  & HEX  & $3$  & $1.536$\\
 $163$  & $46$  & $53.10487$  & $-27.70522$  & $53.10483$  & $-27.70525$  & $0.2$  & 3.3($\pm0.4$)E-15  & $0.08^{-0.05}_{+0.06}$  & $500\pm40$  & 5.2E43  & 5.7E43  & 5.9E21  & QSO-1  & BLAGN  & $3$  & $1.617$\\
 $177$  & $42a$ & $53.11250$  & $-27.68475$  & $53.11250$  & $-27.68475$  & $0.0$  & 6.9($\pm0.2$)E-14  & $0.14^{-0.01}_{+0.01}$  & $910\pm70$  & 1.4E44  & 1.6E44  & 4.3E21  & QSO-1  & BLAGN  & $3$  & $0.734$\\
 $179$  & $41$  & $53.11504$  & $-27.69583$  & $53.11508$  & $-27.69583$  & $0.1$  & 1.2($\pm0.1$)E-14  & $1.38^{-0.06}_{+0.07}$  & $390\pm10$  & 6.9E42  & 2.2E43  & 1.2E23  & AGN-2  & HEX  & $3$  & $0.668$\\
 $188$  & $202$ & $53.12441$  & $-27.85163$  & $53.12441$  & $-27.85161$  & $0.1$  & 3.2($\pm0.4$)E-15  & $1.02^{-0.08}_{+0.10}$  & $86\pm11$   & 3.2E43  & 4.1E44  & 1.0E24  & QSO-2  & HEX  & $3$  & $3.700$\\
 $191$  & $39$  & $53.12491$  & $-27.75827$  & $53.12525$  & $-27.75852$  & $1.4$  & 1.3($\pm0.1$)E-14  & $0.22^{-0.02}_{+0.03}$  & $440\pm60$  & 8.3E43  & 1.0E44  & 1.3E22  & QSO-1  & BLAGN  & $3$  & $1.218$\\
 $193$  & $78$  & $53.12525$  & $-27.75652$  & $53.12525$  & $-27.75655$  & $0.1$  & 2.2($\pm0.3$)E-15  & $0.08^{-0.06}_{+0.07}$  & $120\pm10$  & 9.4E42  & 1.0E43  & 2.6E21  & AGN-1  & BLAGN  & $3$  & $0.960$\\
 $195$  & $38$  & $53.12591$  & $-27.75125$  & $53.12625$  & $-27.75150$  & $1.4$  & 7.2($\pm0.5$)E-15  & $0.09^{-0.03}_{+0.03}$  & $175\pm30$  & 1.6E43  & 1.7E43  & 2.5E21  & AGN-1  & BLAGN  & $3$  & $0.738$\\
 $197$  & $563$ & $53.13112$  & $-27.77305$  & $53.13141$  & $-27.77350$  & $1.9$  & $<$3.3E-16       & $<0.71$                 & $370\pm110$ & $<6.1$E42 & $<1.2$E43 & $<2.5$E23  & AGN-1  & HEX  & $3$  & $2.223$\\
 $214$  & $34a$ & $53.14558$  & $-27.91972$  & $53.14566$  & $-27.91977$  & $0.3$  & 3.6($\pm0.4$)E-15  & $0.38^{-0.06}_{+0.06}$  & $230\pm10$  & 8.4E42  & 1.2E43  & 1.9E22  & AGN-1  & HEX  & $3$  & $0.839$\\
 $219$  & $901$ & $53.14883$  & $-27.82111$  & $53.14883$  & $-27.82111$  & $0.0$  & 7.0($\pm2$)E-16    & $1.11^{-0.19}_{+0.34}$  & $610\pm10$  & 3.8E42  & 3.8E43  & 8.3E23  & AGN-1  & HEX  & $3$  & $2.578$\\
 $229$  & $32$  & $53.15600$  & $-27.66680$  & $53.15612$  & $-27.66675$  & $0.5$  & 2.4($\pm0.8$)E-15  & $0.12^{-0.12}_{+0.17}$  & $140\pm30$  & 4.2E42  & 4.5E42  & 3.1E21  & AGN-1  & BLAGN  & $3$  & $0.664$\\
 $230$  & $31$  & $53.15737$  & $-27.87011$  & $53.15741$  & $-27.87011$  & $0.1$  & 8.9($\pm0.5$)E-15  & $0.17^{-0.03}_{+0.03}$  & $1150\pm60$ & 1.1E44  & 1.5E44  & 1.4E22  & QSO-1  & HEX  & $3$  & $1.603$\\
 $234$  & $30a$ & $53.15875$  & $-27.66261$  & $53.15887$  & $-27.66250$  & $0.6$  & 1.6($\pm0.2$)E-14  & $-0.01^{-0.06}_{+0.08}$ & $560\pm50$  & 5.3E43  & 5.1E43  & 5.8E22  & QSO-1  & BLAGN  & $3$  & $0.837$\\
 $241$  & $201b$& $53.16275$  & $-27.74419$  & $53.16225$  & $-27.74427$  & $1.6$  & 2.1($\pm0.3$)E-15  & $0.66^{-0.08}_{+0.10}$  & $<80$       & 2.4E42  &  4.2E42 & 3.8E22  & AGN-2  & HEX  & $3$  & $0.679$\\
 $242$  & $28$  & $53.16283$  & $-27.76713$  & $53.16287$  & $-27.76722$  & $0.3$  & 3.1($\pm0.4$)E-15  & $0.58^{-0.07}_{+0.08}$  & $250\pm30$  & 1.2E43  & 2.5E43  & 5.8E22  & AGN-1  & BLAGN  & $3$  & $1.216$\\
 $245$  & $27$  & $53.16533$  & $-27.81408$  & $53.16529$  & $-27.81402$  & $0.2$  & 7.1($\pm0.6$)E-15  & $0.97^{-0.05}_{+0.06}$  & $150\pm30$  & 6.4E43  & 5.8E44  & 7.6E23  & QSO-2  & HEX  & $3$  & $3.064$\\
 $251$  & $24$  & $53.17445$  & $-27.86736$  & $53.17441$  & $-27.86738$  & $0.2$  & 3.8($\pm0.5$)E-15  & $0.36^{-0.06}_{+0.06}$  & $110\pm30$  & 2.0E44  & 4.6E44  & 1.8E23  & QSO-1  & BLAGN  & $3$  & $3.610$\\
 $254$  & $91$  & $53.17850$  & $-27.78400$  & $53.17850$  & $-27.78405$  & $0.2$  & 1.6($\pm0.3$)E-15  & $0.41^{-0.08}_{+0.09}$  & $<80$       & 5.8E43  & 1.5E44  & 1.9E23  & QSO-1  & BLAGN  & $1$  & $3.193$\\
 $261$  & $21$  & $53.18458$  & $-27.88086$  & $53.18466$  & $-27.88091$  & $0.3$  & 6.8($\pm2$)E-16    & $0.10^{-0.12}_{+0.18}$  & $560\pm100$ & 6.1E43  & 7.6E43  & 3.3E22  & QSO-1  & BLAGN  & $3$  & $3.471$\\
 $275$  & $19$  & $53.19945$  & $-27.69663$  & $53.19966$  & $-27.69666$  & $0.7$  & 1.3($\pm0.1$)E-14  & $0.19^{-0.04}_{+0.05}$  & $230\pm35$  & 2.7E43  & 3.1E43  & 6.2E21  & AGN-1  & BLAGN  & $3$  & $0.733$\\
 $278$  & $268a$& $53.20512$  & $-27.68050$  & $53.20500$  & $-27.68072$  & $0.9$  & 8.3($\pm2$)E-15    & $>1.36$ & $2360\pm600$                 & 1.2E43  & 7.1E43  & $>2.7$E23  & AGN-2  & HEX  & $3$  & $1.222$\\
 $286$  & $15$  & $53.22029$  & $-27.85547$  & $53.22033$  & $-27.85555$  & $0.3$  & 5.4($\pm0.8$)E-15  & $0.27^{-0.04}_{+0.05}$  & $140\pm50$  & 3.3E43  & 4.6E43  & 1.9E22  & AGN-1  & BLAGN  & $1$  & $1.227$\\
 $290$  & $101a$& $53.23116$  & $-27.79763$  & $53.23125$  & $-27.79775$  & $0.5$  & 1.3($\pm0.3$)E-15  & $0.27^{-0.09}_{+0.12}$  & $80\pm20$   & 1.5E43  & 2.2E43  & 2.7E22  & AGN-1  & BLAGN  & $3$  & $1.625$\\
 $301$  & $13$  & $53.24595$  & $-27.72766$  & $53.24620$  & $-27.72763$  & $0.8$  & 8.7($\pm0.8$)E-15  & $0.22^{-0.04}_{+0.05}$  & $270\pm40$  & 1.7E43  & 2.0E43  & 7.2E21  & AGN-1  & BLAGN  & $3$  & $0.733$\\
 $305$  & $11$  & $53.24929$  & $-27.79669$  & $53.24929$  & $-27.79672$  & $0.1$  & 1.1($\pm0.6$)E-14  & $0.19^{-0.03}_{+0.03}$  & $100\pm30$  & 4.3E44  & 6.2E44  & 4.0E22  & QSO-1  & BLAGN  & $3$  & $2.579$\\
 $311$  & $77$  & $53.25641$  & $-27.76175$  & $53.25641$  & $-27.76183$  & $0.3$  & 1.8($\pm0.4$)E-15  & $0.13^{-0.09}_{+0.11}$  & $380\pm20$  & 2.6E42  & 2.8E42  & 3.0E21  & AGN-1  & BLAGN  & $3$  & $0.622$\\
 $316$  & $4a$  & $53.26500$  & $-27.75513$  & $53.26512$  & $-27.75525$  & $0.6$  & 6.4($\pm0.6$)E-15  & $0.29^{-0.04}_{+0.05}$  & $<80$       & 4.1E43  & 5.8E43  & 2.0E22  & AGN-1  & BLAGN  & $1$  & $1.260$\\
\enddata
\tablecomments{Columns are as in table~\ref{tab:sample}.}
\end{deluxetable}
\clearpage
\end{landscape}
\setlength{\hoffset}{0mm}
\begin{deluxetable}{lllll}
\tabletypesize{\tiny}
\tablecolumns{5}
\tablecaption{Axis Ratios for the Optically--Dull Sample.\label{ba_optdull}}
\tablehead{
\colhead{AID}  &  \colhead{Best Model} & \colhead {galfit b/a} & \colhead {sextr b/a} 
& \colhead{morphology notes}\\
}
\startdata 
%
\cutinhead{AGN with redshifts $z<0.5$ or $z>0.8$}
65  &   34 &  0.24 & 0.34 & dusty edge-on galaxy\\ 
84  &   -  &  int  & 0.60 & small antennae\\
90  &   4  &  irr  & ---  & irregular\\    
131 &   1  &   pt  & 0.95 & pt source\\     
162 &   4  &  0.87 & 0.77 & compact\\ 
164 &   2  &  0.74 & 0.76 & face-on disk +br bulge\\     
166 &   1  &  pt   & 0.35 & faint pt. src\\
171 &   2  &  irr  & 0.51 & irr with central bulge\\
216 &   34 &  0.63 & 0.49 & compact w fuzz\\   
221 &   4  &  0.68 & 0.77 & compact w fuzz\\           
259 &   -  &  irr  & 0.80 & irregular\\  
264 &   1  &  pt   & 0.86 & compact w fuzz\\             
271 &   4  &  0.62 & 0.59 & bulge + fuzz\\            
303 &   3  &  0.33 & 0.37 & edge-on spiral\\             
\cutinhead{AGN with redshifts $0.5<z<0.8$}
44   &  4  &  0.43 & 0.69 & edge-on bulge+disk galaxy\\
48   &  -  &  irr  & 0.64 & clumpy, v. faint\\
60   &  -  &  irr  & 0.38 & compact src + fuzz\\          
80   &  4  &  0.84 & 0.88 & pt + fuzz\\              
83   &  24 &  0.87 & 0.86 & compact but extended\\
91   &  2  &  0.34 & 0.43 & large clumpy spiral\\         
126  &  13 &   pt  & 0.86 & compact w some faint emiss\\
129  &  4  &  $<0.61$ & 0.94 & edge-on Sp w bright bulge\\
134  &  2  &  0.32 & 0.47 & edge-on sp w bright bulge\\
139  &  4  &  0.68 & 0.65 & face-on spiral w bright bulge?\\
146  &  2  &  0.26 & 0.41 & edge-on sp w bright bulge\\  
155  &  2  &  0.73 & 0.75 & irr w bright bulge\\    
157  &  4  &  0.50 & 0.57 & somewhat inclined disk\\     
161  &  4  &  0.85 & 0.90 & compact\\   
176  &  4  &  0.68 & 0.66 & face-on Sp w central bulge\\ 
196  &  -  &  int  & 0.72 & merger/interaction\\
212  &  24 &  0.89 & 0.84 & compact\\        
220  &  23 &  0.63 & 0.68 & compact w fuzz\\             
227  &  4  &  0.61 &\nodata& compact w fuzz\\           
247  &  24 &  0.74 & 0.89 & compact w faint fuzz\\        
269  &  24 &  0.60 & 0.60 & bright bulge +disk\\
276  &  23 &  0.78 & 0.75 & compact+fuzz\\
\enddata
\tablecomments{Columns:  (1) X-ray source identification from \citet{alex}.  
Columns:  (1) X-ray source identification from \citet{alex}.  
(2)  Best-fitting Galfit model (1=PSF, 2=Sersic, 3=PSF+Sersic, 4=bulge+disk).
(3)  Axis ratio from best-fitting Galfit model.
(4)  Axis ratio from best-fitting Source Extractor model.
(5)  Notes on galaxy morphology.
}
\end{deluxetable}
\clearpage
\begin{deluxetable}{lllll}
\tabletypesize{\tiny}
\tablecolumns{5}
\tablecaption{Axis Ratios for the Optically--Active Sample.\label{ba_optact}}
\tablehead{
\colhead{AID}  &  \colhead{Best Model} & \colhead {galfit b/a} & \colhead {sextr b/a} 
& \colhead{morphology notes}\\
}
\startdata  
%
\cutinhead{AGN with $z<0.5$ or $z>0.8$}
 34 & 1  &  pt  & 0.99 &  slightly resolved\\
 39 & 1  &  pt  & 0.93  &  point source\\
 41 & 1  &  pt  & 0.98  &  point source + faint compact fuzz\\
 43 & 1  &  pt  & 0.96  &  point source \\
 68 & 1  &  pt  & 0.98  &  point source\\
 76 & 1  &  pt  & 0.94  &  point source + faint compact fuzz\\
 86 & -  &  --  & 0.56  &  small interm spiral or irr\\  
 94 & 4  &  0.97 & 0.82 & irr. intereaction? \\
 96 & 4  &  irr & 0.74  &  irregular\\
122 & 1  &  pt  & 0.91  &  point source\\
123 & 4  &  0.94& 0.85  & compact src + fuzz\\
163 & 1  &  pt  & 0.87  &  compact src\\
188 & 1  &  pt  & 0.83  &  point source \\
191 & 4  &  0.54& 0.50  & compact src + round fuzz\\
193 & -  &  int & 0.97  &   2 bright peaks + fuzz,  merger?\\
197 & -  &  irr &\nodata&  irr : bright core + offset fluff\\
214 & -  &  **  & 0.69  &  face-on sp or interaction?\\
219 & 1  &  pt  & 0.49  &  faint point source + fuzz\\
230 & 1  &  pt  & 0.92  &  point source + faint compact fuzz\\
234 & 14 &  pt  & 0.89  &  point source + faint fuzz\\
242 & 1  &  pt  & 0.87  &  point source + fuzz\\
251 & 1  &  pt  & 0.99  &  point source\\
254 & 1  &  pt  & 0.91  &  point source\\
261 & 1  &  pt  & 0.96  &  point source\\
286 & 14 &  0.59\tablenotemark{a} & 0.70  & psf + outer ring\\
%
%
\cutinhead{AGN with $0.5<z<0.8$}
66  & 3 &  0.86 & 0.92 & point source + face-on disk\\
88  & - &  ---  & 0.62 & interacting face-on spiral?\\
103 & 4 &  0.86 & 0.93 & round halo + possible inclined disk\\ 
117 & - &  0.88\tablenotemark{a} & 0.79 & compact, round\\
177 & 3 &  0.90 & 0.93 & psf + face-on disk\\
179 & 3 &  0.79 & 0.81 & compact   \\
195 & - &  0.85\tablenotemark{a} &\nodata& compact, round\\
229 & 4 &  0.88 & 0.90 & bright center + fuzz\\
241 & 1 &  --   & 0.73 & compact\\
\enddata
\tablecomments{Columns:  (1) X-ray source identification from \citet{alex}.  
Columns:  (1) X-ray source identification from \citet{alex}.  
(2)  Best-fitting Galfit model (1=PSF, 2=Sersic, 3=PSF+Sersic, 4=bulge+disk).
(3)  Axis ratio from best-fitting Galfit model.
(4)  Axis ratio from best-fitting Source Extractor model.
(5)  Notes on galaxy morphology.
}
\tablenotetext{a}{Galfit crashed, so the axis ratio was fit using IRAF's task {\bf ellipse.}}

\end{deluxetable}

\clearpage
\appendix
\section{Appendix:  K-correcting model 24~\micron/H$\beta$ ratios}
\label{sect:appendix}

We predict how K-corrections should affect the observed 
$f_{\nu}(24~\micron)/f(H\beta)$ ratio, as follows.  
We redshift each \citet{charyelbaz} model to find the 
observed 24~\micron\ flux density normalized to the rest--frame 
15~\micron\ flux density; 
this is a K-correction and thus depends on redshift and SED shape.
We normalize to 15~\micron\ in order to take advantage of the
empirical relation between 15~\micron\ and Balmer emission
measured by \citet{roussel}, which, combining equations from 
\citet{roussel} and \citet{rob98}, we can write as follows:
\begin{equation}
\bigg( \frac{f_{\nu}(15~\micron)^{rest}}{f(H\beta)^{rest}} \bigg) = 
\bigg( \frac{14 \times 10^{26}}{\Delta \nu_{ISO}} \bigg)
\frac{mJy}{erg~s^{-1}~cm^{-2}}
\end{equation}

where $\Delta \nu_{ISO} = 6.75 \times 10^{12}$~Hz. 

We can now combine this 15~\micron--Balmer relation with the 
K-correction to calculate how the observed $f_{\nu}(24~\micron)/f(H\beta)$ 
ratio changes as the source is moved to higher redshift.
\begin{equation}
\bigg( \frac{f_{\nu}(24~\micron)^{obs}}{f(H\beta)^{obs}} \bigg)   = 
\bigg( \frac{f_{\nu}(15~\micron)^{rest}}{f(H\beta)^{rest}} \bigg)
\bigg( \frac{f_{\nu}(24~\micron)^{obs}}{f_{\nu}(15~\micron)^{rest}} \bigg)
\end{equation}

In figure~\ref{fig:hbeta} we plot this result, the expected 
$f_{\nu}(24~\micron)/f(H\beta)$ ratios for \citet{charyelbaz} models.

\end{document}